\documentclass[12pt]{article}
\usepackage{amsmath}
\usepackage{amssymb}
\begin{document}
\begin{center}
{ {\large   {(Anti-)Chiral Supervariable Approach to Nilpotent and Absolutely Anticommuting Conserved Charges of Reparameterization Invariant Theories: A Couple of Relativistic Toy Models as Examples }}}

\vskip 3.0cm

{\sf S. Kumar$^{(a)}$, B. Chauhan$^{(a)}$, R. P. Malik$^{(a,b)}$}\\
$^{(a)}$ {\it Physics Department, Center of Advance Studies, Institute of Science,}\\
{\it Banaras Hindu University, Varanasi - 221 005, (U.P.), India}\\

\vskip 0.1cm

\vskip 0.1cm

$^{(b)}$ {\it DST Center for Interdisciplinary Mathematical Sciences,}\\
{\it Institute of Science, Banaras Hindu University, Varanasi - 221 005, India}\\
{\small {\sf {e-mails: sunil.bhu93@gmail.com; bchauhan501@gmail.com;  rpmalik1995@gmail.com}}}

\end{center}

\vskip 2cm

\noindent
{\bf Abstract:} We exploit the potential and power of the   Becchi-Rouet-Stora-Tyutin (BRST) and
anti-BRST invariant 
restrictions on the (anti-)chiral supervariables to derive the proper nilpotent (anti-)BRST
 symmetries for the reparameterization
invariant one (0+1)-dimensional (1D) toy models of a free relativistic
 particle as well as a free spinning (i.e. supersymmetric)
relativistic  particle  within the framework of  (anti-)chiral supervariable approach 
to BRST formalism. Despite the (anti-)chiral super expansions of the (anti-)chiral supervariables, we observe that
the (anti-)BRST charges, for the above toy models, turn out to be absolutely anticommuting in nature. This is one of the novel
observations of our present endeavor. For this proof, we utilize the beauty and strength  of Curci-Ferrari (CF)-type restriction
in the context of a spinning relativistic particle but {\it no} such  restriction is required in the case of a free scalar relativistic particle.
We have also captured the nilpotency property of the conserved charges as well as the (anti-)BRST invariance of the appropriate
 Lagrangian(s) of our present toy models within the framework of (anti-)chiral supervariable approach.

 \vskip 1.5cm


\vskip 0.25cm


\newpage
\section{Introduction}

The usual superfield approach (USFA) to Becchi-Rouet-Stora-Tyutin (BRST) formalism [1-8] exploits the idea of horizontality condition (HC) where 
the concepts from differential geometry play a decisive role. This approach enables us to derive the (anti-)BRST symmetries 
associated with the $p$-form ($p$ = 1,2,3,...) gauge field and associated (anti-)ghost fields of a given $p$-form gauge theory
within the framework of BRST formalism. In addition, it also leads to the derivation of Curci-Ferrari (CF) condition [9] 
which is the hallmark [10,11] of a given {\it quantum} gauge theory  within the framework of  BRST formalism. Moreover,
the USFA (with the help of the geometrical HC) {\it also} sheds light on the {\it geometrical meaning } of the nilpotency and absolute
anticommutativity properties of the 
BRST and anti-BRST symmetries and corresponding conserved charges (which are primarily  mathematical in nature
{\it without} the knowledge of USFA to BRST formalism). These observations are true for a given {\it free} $p$-form 
gauge theory (in any arbitrary dimension of spacetime) where there is {\it no} interaction between the gauge field and matter field(s).

The above USFA has been systematically {\it generalized} in our earlier works [12-16] where we have exploited 
the theoretical strength of gauge invariant restrictions (along {\it and} consistent with HC) so as to derive the
(anti-)BRST symmetries associated with the {\it matter}, gauge and (anti-)ghost fields {\it together}. This generalized 
version of superfield approach to BRST formalism has been christened  as the augmented superfield approach (ASFA) to BRST formalism
in our earlier works [12-16]. In our present endeavor, we shall exploit the theoretical potential and power of 
ASFA\footnote{To be precise, we shall utilize the augmented version of the {\it supervariable} approach
to BRST formalism where we shall consider only the (anti-)chiral super expansions of the  supervariables. We christen
our approach as `` the supervariable approach'' because we are dealing with ``variables'' and {\it not} the ``fields'' in our present endeavor
(where we are concerned with {\it only} the 1D toy models of relativistic particles).} 
to BRST formalism
to capture the nilpotency and absolute anticommutativity properties of the (anti-)BRST charges for the reparameterization 
invariant theories of a 1D free massive {\it scalar }  and a massless {\it spinning} relativistic particles. 
The clinching proof of the absolute anticommutativity
property of the conserved and nilpotent (anti-)BRST charges is a {\it novel} observation in our present endeavor.

In all the above superfield approaches [1-8, 12-16], we have taken super expansions of the superfields along {\it all} possible Grassmannian
directions of the appropriately chosen (D, 2)-dimensional supermanifold on which a given D-dimensional $p$-form gauge theory is generalized.
In our recent set of papers [17-22], we have taken the help of (anti-)chiral  supervariables/superfields 
to derive the (anti-)BRST as well as (anti-)co-BRST symmetry transformations for the 1D toy model of a rigid rotor [17, 18], 2D model of a 
self-dual bosonic field theory [19] and 4D model of an Abelian 2-form gauge theory [20]. To derive the above symmetries, we have utilized 
the idea of symmetry  invariance where we have invoked the  (anti-)BRST and (anti-)co-BRST invariant  restrictions on the supervariables/superfields.
We have also applied the above idea in the context of {\it interacting} Abelian and non-Abelian gauge theories where there are  
interactions between the gauge  and matter fields [21, 22]. One of the novel observations of our earlier endeavors [17-22]
has been the proof of  absolute anticommutativity of the (anti-)BRST (and (anti-)co-BRST) charges {\it despite} the
fact that we have considered  {\it only} the (anti-)chiral super expansions of the supervariables/superfields. This observation has been established
{\it only} in the context of {\it gauge theories} within the framework of (anti-)chiral superfield approach to BRST formalism.

The central theme of our present investigation is to exploit the simplicity and beauty of the  augmented
version of (anti-)chiral supervariable approach [(A)CSVA] to BRST formalism\footnote{We call our approach as
 the {\it supervariable} approach to BRST formalism because, in the limiting case (when the Grassmannian coordinates are set equal to zero), we
obtain a {\it variable} from the super expansion(s) of the supervariable(s). 
This observation should be contrasted with the {\it superfield} approach to BRST formalism where we obtain
a {\it field}, in the {\it above} limit, from the super expansion(s) of the superfield(s).} in the context of  
reparameterization invariant theories of 1D toy models of a free {\it scalar}  relativistic particle as well as a {\it spinning} relativistic 
particle to establish the absolute anticommutativity property of the (anti-)BRST charges. In the process,
we {\it also} capture the nilpotency of the (anti-)BRST charges and the (anti-)BRST invariance of the appropriate 
Lagrangian(s) of the above models.  The present reparameterization invariant theories are important because these are 
precursors to the (super)string  and supergravity theories where the idea of reparameterization invariance
plays a key role. It is worthwhile to point out that the {\it gauge} and {\it reparameterization} 
symmetries of the present toy models have been shown to be {\it equivalent} when the free motion ($\dot p_\mu = 0$)
of the free relativistic particles and the specific relationship between the gauge and reparameterization transformation parameters  
are taken into account {\it together} (see. e.g. [24] for details).

We would like to lay emphasis that, in the proof of  absolute anticommutativity of the 
(anti-)BRST charges, we have utilized the beauty and strength of CF-type restriction which has been systematically and elegantly
derived in our earlier work in the case of a 1D toy model of a {\it spinning} relativistic particle [23].
However, there is {\it no} CF-type restriction in the context of a free massive {\it scalar} relativistic 
particle. As a consequence, for the proof of the 
absolute anticommutativity of the conserved (anti-)BRST charges (in the case of {\it this } 1D toy model), there is {\it no} 
requirement  of invoking any kind of {\it restriction} from {\it  outside}. Our present endeavor {\it completes} our program 
of proving the nilpotency and absolute anticommutativity of the (anti-)BRST charges within the 
framework of (A)CSVA/(A)CSFA to {\it gauge invariant}
theories and {\it reparameterization invariant} theories (where {\it only} the (anti-) chiral super expansions of 
the supervariables/superfields  are utilized).
We re-emphasize that, for the 1D toy models under consideration in our present endeavor, the gauge and reparameterization symmetries are
intertwined together in a beautiful manner and they are found to be {\it equivalent} on-shell under very 
{\it specific} condition where the transformation parameters 
of these symmetry transformations are related with each-other in a particular fashion (see, e.g. [24] for details).

We enumerate a few novel features that are associated with the discussion of a
 free $spinning$ relativistic particle (against the backdrop
of our discussions connected with a free $scalar$ relativistic particle)
 within the framework of (A)CSVA to BRST formalism. The conserved (anti-)BRST 
charges (cf. Eq. (4) below), derived directly from the application of 
the Noether theorem, are found to be {\it off-shell} nilpotent in the case of a free $scalar$ 
relativistic particle. However, the same charges 
$Q_{(a)b}^{(1)}$ (cf. Eq. (13) below), derived directly by exploiting the virtues of Noether's theorem,
turn out to be {\it on-shell} nilpotent (despite the fact that we have used the (anti-)BRST
 symmetries (cf. Eqs. (9), (10) below) which are off-shell nilpotent).
These charges become off-shell nilpotent {\it only} when we use 
the appropriate EOMs (cf. Eqs. (14), (15) below) to recast them in a different form
$Q_{(a)b}^{(2)}$ (cf. Eqs. (13), (17) below). In exactly similar fashion, the
 (anti-)BRST charges (cf. Eq. (4) below) are absolutely anticommuting {\it without}
any use of EOMs and/or any kind of outside restriction(s)
 (e.g. CF-type condition(s)). This is not the case with the (anti-)BRST charges associated
with the {\it spinning} relativistic particle under consideration. 
In the proof of absolute anticommutativity property of the (anti-)BRST charges for
a {\it spinning} relativistic particle, we have to {\it first} 
recast the expressions for the off-shell nilpotent charges (cf. Eq. (17) below) in a different form
(cf. Eq. (63) below) by utilizing the appropriate EOMs as well 
as the CF-type restriction and, then only, we have to exploit the idea of continuous symmetries
and their generators to prove the absolute 
anticommutativity property of the (anti-)BRST charges (cf. Eqs. (66), (70) below).

Our present investigation is essential and interesting 
on the following  grounds. First, we have applied, so far, our (A)CSVA/(A)CSFA
to {\it only} 1D, 2D and 4D models of {\it gauge theories} and derived the (anti-)BRST
 as well as (anti-)co-BRST symmetries.
Thus, it is of immense  importance for us to apply {\it it} to the models of 
reparametrization invariant theories (which are the precursors 
to the (super)string and supergravity theories). 
Second, the results of our present investigation establish that the absolute
 anticommutativity
of the (anti-)BRST charges is {\it universal} in the 
case of gauge- and reparameterization invariant theories {\it despite}
the fact that we have taken into account {\it only} the (anti-)chiral super 
expansions for the supervariables/superfields within the framework
of (A)CSVA/(A)CSFA to BRST formalism. Finally, the reparameterization 
invariant models (under consideration) are 
interesting and  important in their own right as they {\it also} 
represent the {{\it ordinary} as well as the {\it supersymmetric} 
prototype toy models of theoretical (and mathematical) interests in one (0+1)-dimension of spacetime.

Our present paper is organized as follows. First of all, 
to fix the notations and convention, we discuss
 concisely, in Sec. 2,  the bare essentials of a free scalar
 relativistic and a free spinning relativistic particles within the 
framework of (anti-) BRST invariant Lagrangians where we elaborate on the 
nilpotent (anti-)BRST symmetries  and derive their corresponding
 conserved charges. Our Sec. 3 is devoted to the derivation of
(anti-)BRST symmetry transformations for a free scalar relativistic 
particle within the framework of (A)CSVA to BRST formalism
where we invoke the (anti-)BRST invariant restrictions on the
 (anti-)chiral supervariables. In Sec. 4, we derive the 
nilpotent (anti-)BRST symmetry transformations for a spinning 
relativistic particle by exploiting the beauty and strength
of the (anti-)BRST invariant restrictions on the (anti-)chiral 
supervariables (that are fermionic, bosonic 
and their appropriate  combinations). Our Sec. 5 deals with the 
proof of nilpotency and absolute anticommutativity of the 
(anti-)BRST charges for {\it both} of our models within the
 framework of (A)CSVA to BRST formalism. Finally,  we summarize 
 our key results, comment on some crucial issues related to our 
1D reparameterization invariant\footnote{It can be checked that the infinitesimal 
version of reparameterization transformation leaves the action 
integrals, corresponding  to the appropriate Lagrangian(s) of our 1D  toy models, invariant 
(see, e.g. Refs. [23-25] for details).}
 toy models   and point out a few future directions for further investigations in our last section (i.e. Sec. 6).

In our Appendices A, B and C, we elaborate on a few theoretical
 computations that have  {\it either} been incorporated into the 
main body of our text {\it or} they supplement our key results.
 Particularly, in our Appendix C, we capture the (anti-)BRST
invariance of the Lagrangian(s) of the {\it two} relativistic toy models (existing in one (0+1)-dimensional spacetime)
 that have been considered in our present endeavor.\\

\noindent
{\it Convention and Notations}: Through out the whole
 body of our text, we take the convention of left-derivative w.r.t. all
the fermionic variables (e.g. $c, \bar c, \chi, \psi_\mu, \gamma$) 
for all the appropriate computations that are connected with
the 1D toy models of our present endeavor.
We also use the notations $s_{(a)b}$  and $Q_{(a)b}$ for the
 off-shell nilpotent (anti-)BRST symmetry transformations 
and corresponding off-shell nilpotent (anti-)BRST charges for {\it both} the relativistic toy
models which are examples of the reparameterization invariant theories.

\section{Preliminary: Off-Shell Nilpotent BRST and Anti-BRST Symmetries in Lagrangian Formulation}

We  begin with the following (anti-)BRST invariant 
first-order Lagrangian for a free $scalar$ relativistic particle 
 of  rest mass $m $ (see, e.g. [24, 25]):
\begin{eqnarray}
&&L_{b} =   p_\mu \; \dot x^\mu - \frac{1}{2}\;e\; (p^2 - m^2) + b\; \dot e 
+ \frac{1}{2}\; b^2 - i\; \dot {\bar c}\; \dot c,
\end{eqnarray} 
where the canonically conjugate target space variables $(x_\mu(\tau ), p^\mu(\tau))$ are the D-dimensional 
(i.e. $\mu,\nu,\lambda,... = 0, 1, 2,...,D-1)$ coordinates and momenta, $e(\tau)$ is an einbein variable,
$b(\tau)$ is the Nakanishi-Lautrup auxiliary variable and (anti-)ghost variables ($\bar c(\tau))c(\tau)$ are
needed for the unitarity in the theory. The above Lagrangian describes the {\it free} motion ($\dot p_\mu = 0$)
of a scalar relativistic  particle on a world-line parametrized by $\tau$ and this world-line is embedded in the 
D-dimensional target space. Thus, all the variables of this theory are 
function of $\tau$ and, therefore, we have  $\dot p_\mu  = \frac {d\,p_\mu }{d\,{\tau}}$. It is evident that $\Pi_e \approx 0$ and $p^2 - m^2 \approx 0$
are the first-class constraints on the theory in the terminology of Dirac's prescription for the classification scheme
of constraints [26, 27]. Here $\Pi_e$ is the canonical conjugate momentum w.r.t. the 
einbein variable $e(\tau)$. The above constraints generate the local gauge transformations which can be generalized to the (anti-)BRST symmetry
transformations $s_{(a)b}$ as (see, e.g. [24, 25] for details):
\begin{eqnarray}
s_{ab}\, x_\mu &=& \bar c \;p_\mu, \; s_{ab} \,\bar c = 0, \; 
s_{ab}\, p_\mu = 0, \; s_{ab}\, c = - i\, b, \; s_{ab} b = 0, \;\; s_{ab}\, e = \dot {\bar c},\nonumber\\
s_b\, x_\mu &=& c \,p_\mu, \; s_b\, c = 0, \; s_b \,p_\mu = 0, \;
s_b\, \bar c = i\, b, \; s_b \,b = 0, \; s_b\, e = \dot c.  
\end{eqnarray} 
It is straightforward to check that the above transformations are off-shell nilpotent of order {\it two}  
(i.e. $s_{(a)b}^2 = 0$) and absolutely anticommuting (i.e. $s_b s_{ab} + s_{ab} s_b = 0$) in nature.
We note that, under the above (anti-)BRST symmetry transformations (2), the Lagrangian (1) transforms to the total
derivatives w.r.t. the evolution parameter $\tau $, as  
\begin{eqnarray}
s_b L_b = {\displaystyle \frac{d} {d \tau}} \;
\Bigl [ \frac{1}{2}\; c\; (p^2 + m^2) + b\; \dot c \Bigr ], \qquad\;
s_{ab} L_b = {\displaystyle \frac{d} {d \tau}} \;
\Bigl [ \frac{1}{2}\; \bar c\; (p^2 + m^2) + b\; \dot {\bar c} \Big ],
\end{eqnarray} 
thereby rendering the action integral $S = \int d\tau L_b$ (corresponding to the starting Lagrangian $L_b$) invariant 
 for the physically well-defined variables that vanish-off at $\tau = \pm \infty$ due to Gauss's divergence  theorem.

According to Noether's theorem, the above continuous symmetries lead to the derivation of conserved $({\dot Q_{(a)b}} = 0)$
(anti-)BRST  charges $Q_{(a)b}$ as 
\begin{eqnarray}
Q_b  = \frac {c}{2}\; (p^2-m^2) + b\,\dot c\equiv b\,\dot c- \dot b\,c,\;\;\quad
Q_{ab}  = \frac {\bar c}{2}\; (p^2-m^2) + b\,\dot {\bar c}\equiv b\,\dot {\bar c}- \dot b\,{\bar c},
\end{eqnarray}    
where we have used the equation of motion w.r.t. $e$ which leads to:
$\dot b = -\;\frac {1}{2}\,(p^2-m^2)$ in the {\it final} expressions for $Q_{(a)b}$.
The conserved charges $Q_{(a)b}$ are the generators of the (anti-) BRST symmetry transformations in (2).
The nilpotency and absolute anticommutativity of the (anti-)BRST charges (cf. Eq. (4)) can be proven in a straightforward
fashion as follows
\begin{eqnarray}
&& s_b\,Q_b  = -i\, {\{Q_b, Q_b}\} = 0,\qquad\qquad\quad\;\; s_{ab}\,Q_b = -\, i \;{\{Q_b, Q_{ab}}\} = 0,\nonumber\\
&&s_{ab}\,Q_{ab} = -i\,{\{ Q_{ab},Q_{ab}}\} = 0, \qquad\qquad s_b\,Q_{ab} = - \,i\; {\{Q_{ab}, Q_b}\} = 0,
\end{eqnarray} 
where we have used the property of $Q_{(a)b}$ as the generators for {\it all} the continuous symmetry transformations 
$s_{(a)b}$. We have also applied the symmetry transformations (2) {\it directly} on  $Q_{(a)b}$ to prove that
$s_b\,Q_b = 0, s_{ab}\,Q_{ab} = 0, s_b\,Q_{ab} = i\, (b\,\dot b-\dot b\, b) = 0$ and $s_{ab} Q_b = -i\,(b\,\dot b-\dot b\, b) = 0$.

Now we dwell a bit on a massless spinning relativistic particle which is 
described by the following (anti-)BRST invariant coupled (but equivalent) Lagrangians (see, e.g. [23])
\begin{eqnarray}
L_B &=& L_0 + b \;\dot e + b\;(b + 2\;\beta\;\bar\beta) - i\; \dot{\bar c}\;(\dot c 
+ 2\;\beta\;\chi) + 2\;i\;\bar\beta \;\dot c\;\chi
\nonumber\\&-& 2\;e\;(\gamma\;\chi + \bar\beta\;\dot\beta) + 2\;\beta\;\gamma\;\bar c 
+ \bar\beta^2\;\beta^2 + 2\;\bar\beta\; c\;\gamma,\nonumber\\
L_{\bar B} &=& L_0  - \bar b \;\dot e + \bar b\;(\bar b + 2\;\bar\beta\;\beta)
- i\; \dot{\bar c}\;(\dot c + 2\;\beta\;\chi) + 2\;i\;\bar\beta \;\dot c\;\chi
\nonumber\\ &-& 2\;e\;(\gamma\;\chi - \beta\;\dot{\bar\beta}) + 2\;\beta\;\gamma\;\bar c 
+ \bar\beta^2\;\beta^2 + 2\;\bar\beta\; c\;\gamma,
\end{eqnarray}
where 
$L_{0}$ is the first-order Lagrangian for the 1D toy model of a free massless spinning  relativistic
particle as follows (see, e.g. [24] for details):
\begin{eqnarray}
L_0 = p_\mu\; \dot x^\mu - \frac{e}{2}\; p^2 + \frac{i}{2}\; \psi_\mu \;\dot \psi^\mu  
+ i \;\chi\; (p_\mu\; \psi^\mu).
\end{eqnarray}
In the above, the constraints $p^2\approx  0$ and $ p_\mu\,\psi^\mu\approx 0$ are the first-class  in the terminology 
of Dirac's prescription for the classification of constraints and these have been incorporated into the above Lagrangian
through the Lagrange multiplier variables $e(\tau)$ and $\chi (\tau)$. The latter variables are the analogs
of the vierbein and Rarita-Schwinger 
(i.e. gravitino)  fields of the 4D supergravity theory. In our present discussion, these variables 
$e(\tau)$ and $\chi(\tau)$ are also the analogs  of gauge fields of the 4D gauge theory. The fermionic variables
$\psi_\mu$ are the superpartners of  $x_\mu$ and they satisfy: $(\psi_\mu)^2 = 0$, $\psi_\mu\,\psi_\nu + \psi_\nu\,\psi_\mu = 0$,
$\chi\;\psi_\mu + \psi_\mu\,\chi = 0$ because $\chi(\tau)$ is {\it also} fermionic in nature and it is the superpartner 
of the einbein variable $e(\tau)$. We point out that the super world-line, traced out by the motion of the spinning massless relativistic particle,
is parameterized by $\tau$ and it is embedded in the D-dimensional target space supermanifold where $\mu, \nu, \lambda, ... = 0, 1, 2,...D -1$.
It is straightforward to conclude that {\it all} the variables of our present toy model are function of the evolution parameter $\tau$
and $\dot x_\mu = \frac { d\, x_\mu}{d\,\tau}, \dot \psi_\mu = \frac {d\,\psi_\mu}{d\,\tau}$.

We observe that {\it both} the above Lagrangians are {\it equivalent} because both of them respect the (anti-)BRST symmetries provided 
we use the Curci-Ferrari (CF) type restriction\footnote{ We have 
considered the supersymmetrization of the horizontality condition in our 
earlier work [23] on a massless as well as a massive spinning relativistic particle and have derived {\it this} specific CF-type restriction
from the superfield/supervariable approach 
to our {\it present} reparameterization invariant theory.}: $ b + \bar b + 2 \,\bar\beta \,\beta = 0$ which emerges from the Euler-Lagrange
equations of motion derived from the above Lagrangians, namely; 
\begin{eqnarray}
b = - \frac{1}{2} \dot e - \bar \beta \beta, \quad \bar b = \frac{1}{2} \dot e - \bar \beta \beta 
\;\;\Longrightarrow \;\; b + \bar b + 2 \,\bar \beta \,\beta = 0.
\end{eqnarray}
We shall consider our BRST and anti-BRST symmetries {\it only} on a hyper world-line (embedded in the D-dimensional target space) where
the above CF-type restriction is satisfied because the absolute anticommutativity (i.e. ${\{s_b, s_{ab}}\} = 0$)  of the
off-shell  nilpotent (i.e. fermionic; $s_{(a)b}^2 = 0$) (anti-)BRST symmetries $s_{(a)b}$
is {\it also} satisfied {\it only} on {\it this}
hyper world-line. For instance, it can be checked 
that  ${\{s_b, s_{ab}}\} \;e = 0$ and ${\{s_b, s_{ab}}\} \; x_\mu = 0 $  {\it only} when we use the CF-type restriction:
 $b + \bar b + 2\,\beta\bar\beta  = 0$ in their proof of the absolute anticommutativity properties (i.e. $s_b\,s_{ab} + s_{ab} s_b = 0$).

In the above coupled Lagrangians, the auxiliary variables $(b, \bar b)$ are the Nakanishi-Lautrup variables, $(\bar c)c$ 
are the fermionic ($ c^2 = \bar c^2 = 0, c \bar c + \bar c c = 0$) (anti-)ghost variables, $(\bar \beta) \beta$ are the bosonic (anti-)ghost
variables, $( e, \chi)$ are the gauge and super-gauge variables and $\gamma$ is a fermionic ($\gamma^2 = 0$) auxiliary variable. It can be checked that, under the following off-shell nilpotent $(s_{(a)b}^2 = 0)$  (anti-)BRST symmetry transformations  $s_{(a)b}$
\begin{eqnarray} 
&& s_{ab}\; x_\mu = {\bar c}\; p_\mu + \bar \beta \;\psi_\mu, \quad\qquad s_{ab}\; e = \dot {\bar c} + 2 \;\bar \beta\; \chi,  
\;\quad\qquad s_{ab} \;\psi_\mu = i \;\bar \beta\; p_\mu,\nonumber\\
&& s_{ab}\; \bar c = - i \;{\bar \beta}^2, \quad s_{ab}\; c = i\; \bar b, \quad s_{ab}\; \bar \beta = 0, 
\;\quad s_{ab} \; \beta = - i\; \gamma, \quad s_{ab}\; p_\mu = 0, \nonumber\\
&& s_{ab} \;\gamma = 0, \qquad\qquad s_{ab}\; \bar b = 0, \quad\qquad s_{ab}\;\chi = i\; \dot {\bar \beta}, 
\quad s_{ab} \; b =  2\; i\; \bar \beta\; \gamma,
\end{eqnarray}
\begin{eqnarray}
&&s_b\; x_\mu = c\;p_\mu + \beta \;\psi_\mu, \quad\qquad s_b\; e = \dot c + 2\;\beta\; \chi,  
\quad\qquad s_b\; \psi_\mu = i\;\beta\; p_\mu,\nonumber\\
&& s_b\;c = - i\; \beta^2, \;\quad s_b \;{\bar c} = i\; b, \;\quad s_b \;\beta = 0, 
\;\quad s_b \;\bar \beta = i \;\gamma, \;\quad s_b\; p_\mu = 0,\nonumber\\
&& s_b \;\gamma = 0, \qquad\quad s_b \;b = 0, \qquad\quad s_b \;\chi = i\; \dot \beta, 
\qquad\quad s_b\; \bar b = - 2\; i\; \beta\; \gamma,
\end{eqnarray} 
the Lagrangians $L_B$ and  $L_{\bar B}$ transform to total derivatives w.r.t. the evolution parameter $\tau$
(which characterizes the super world-line) as:
\begin{eqnarray}
s_b\; L_B = \frac{d}{d\tau}\;\Bigl[\;\frac{1}{2}\;c\; p^2 + \frac{\beta}{2}\;(p \cdot\psi)  
+ b\;(\dot c + 2\; \beta\; \chi) \Bigr],
\end{eqnarray}
\begin{eqnarray}
s_{ab}\; L_{\bar B} = \frac{d}{d\tau}\;\Bigl[\;\frac{1}{2}\;\bar c\; p^2 + \frac{\bar\beta}{2}\;(\;p \cdot\psi)  
- \bar b\;(\dot {\bar c} + 2\; \bar\beta\; \chi)\Bigr].
\end{eqnarray}
As a consequence of the above explicit transformations, it is evident that the corresponding action integrals
(i.e. $S_1 = \int\, d\,\tau L_B$ and $S_2  = \int \, d\,\tau L_{\bar B}$) would remain
invariant under the BRST and anti-BRST symmetry transformations for the physically well-defined variables that vanish-off at $\tau = \pm \infty$.

Invariance of the action, under the  continuous symmetry transformations, leads to the derivation of the Noether conserved
currents and corresponding conserved charges for our 1D system. We have the following equivalent expressions for the
conserved  (anti-)BRST charges (taking into account the EOM: $\beta\,\dot{\bar c} = -\,\frac{1}{2}(p\cdot\psi) + i\,e\,\gamma +\bar\beta\,\dot c $),
namely;
\begin{eqnarray}
&& Q_{ab}^{(1)}   = \frac {1}{2}\,\bar c\, p^2 -\bar b \,\dot{\bar c} + \bar\beta (p\cdot\psi) - \bar \beta^2\, \dot c  - 2 \beta \,\bar\beta^2\,  \,\chi  - 2\bar b\, \bar \beta\,\chi,\nonumber\\
&&Q_{ab}^{(2)}\equiv  \frac {1}{2}\,\bar c\, p^2 - \bar b \,\dot{\bar  c} - \bar \beta\,\beta\,\dot {\bar c} + i\, \bar\beta\,e\,\gamma +\frac {1}{2} \bar\beta\,(p\cdot\psi) - 2 \bar\beta^2\,\beta   \,\chi -2\bar b\,\bar \beta\,\chi,\nonumber\\
&& Q_b^{(1)}   =  \frac {1}{2}\, c\, p^2 + b \,\dot c + \beta (p\cdot\psi) + \beta^2\, \dot{\bar c} + 2 \beta^2\,\bar\beta   \,\chi + 2 b\, \beta\,\chi,\nonumber\\
&& Q_{b}^{(2)} \equiv  \frac {1}{2}\, c\, p^2 + b \,\dot c +   \beta\,\bar\beta\,\dot c + i\, \beta\,e\,\gamma +\frac {1}{2} \beta\,(p\cdot\psi) + 2 \beta^2\,\bar\beta   \,\chi + 2 b\, \beta\,\chi,
\end{eqnarray} 
where, in expressing the equivalent forms of the charges, we have used the following  equations of motion (EOMs) derived from the Lagrangian $L_B,$ 
namely;
\begin{eqnarray}
&&\dot p_\mu = 0, \quad \dot x_\mu = e\,p_\mu - i\,\chi\,\psi_\mu,\quad \dot\psi_\mu = \chi\,p_\mu,\quad \dot e + 2\,\beta\,\bar\beta +  2\,b = 0,\
\nonumber\\
&& \dot b = -\,\frac {p^2}{2} - 2\,\gamma\,\chi - 2\,\bar\beta \dot \beta, \quad e\,\dot {\bar\beta} + \dot e\, \bar\beta + b\,\bar\beta
-\,i\dot{\bar c}\,\chi + \gamma\,\bar c + \bar\beta^2\,\beta = 0,\nonumber\\
&& \beta b + i\,\dot c\,\chi - e\dot \beta + \bar\beta\,\beta^2 + c\,\gamma = 0, \quad 2\,\beta\, \dot {\bar c}  - 2\, {\bar \beta}\,\dot c
- 2\,i\,e\,\gamma + (p\cdot\psi) = 0,\nonumber\\
&&\beta\,\bar c - \bar\beta\,c - e\,\chi = 0,\quad \ddot  {\bar c} + 2\,\dot {\bar \beta}\, \chi + 2\,\bar\beta\,\dot\chi + 2\,i\,\bar\beta\,\gamma = 0,
\nonumber\\
&& \ddot c + 2\,i\beta\,\gamma + 2\,\dot\beta\,\chi + 2\,\beta\,\dot\chi = 0,
\end{eqnarray}
and the EOMs, emerging from the equivalent Lagrangian $L_{\bar B}$ (that are different from the {\it above} EOMs) are
as follows:
\begin{eqnarray}
&&\dot e  - 2\,\beta\,\bar\beta -  2\,\bar b = 0,\quad \quad \quad \bar\beta\bar b 
- i\,\dot {\bar c }\,\chi + e\dot{\bar\beta} + \beta\,\bar\beta^2 - \bar c\,\gamma = 0,
\nonumber\\
&& \dot{\bar b} = \,\frac {p^2}{2} + 2\,\gamma\,\chi - 2\,\beta \dot {\bar\beta}, \quad
e\,\dot \beta + \dot e\, \beta - \bar b\,\beta
-\,i\dot c\,\chi +\gamma\, c - \beta^2\,\bar\beta = 0.
\end{eqnarray}
We note that the expressions for $Q_b^{(1)}$ and $Q_{ab}^{(1)}$ have been obtained from the {\it direct} use of the
Noether theorem related with the continuous symmetries. It can be checked explicitly that the following 
\begin{eqnarray}
&& s_b Q_b  = -\,i\, {\{Q_b, Q_b}\} = 0\quad\qquad\;\; \Longrightarrow\;\;\;\;\; Q_b^2 = 0,\nonumber\\
&& s_{ab} Q_{ab}  = -\,i\, {\{Q_{ab}, Q_{ab}}\} = 0 \quad\quad \Longrightarrow \;\;\;\;\;Q_{ab}^2 = 0,\nonumber\\
&& s_b Q_{ab}  = -\,i\, {\{Q_{ab}, Q_b}\} = 0 \;\;\, \Longrightarrow \;\;\;\;\; s_{ab} Q_b  = -\,i\, {\{Q_b, Q_{ab}}\} = 0,
\end{eqnarray}
are true if we choose the specific set  of the above (anti-)BRST charges, namely;
\begin{eqnarray}
&&Q_b \equiv  Q_b^{(2)} =  \frac {1}{2}\, c\, p^2 + b \,\dot c +   \beta\,\bar\beta\,\dot c + i\, \beta\,e\,\gamma +\frac {1}{2} \beta\,(p\cdot\psi) + 2 \beta^2\,\bar\beta   \,\chi + 2 b\, \beta\,\chi,\nonumber\\
&& Q_{ab}\equiv Q_{ab}^{(2)} =   \frac {1}{2}\,\bar c\, p^2 - \bar b \,\dot{\bar  c} - \bar \beta\,\beta\,\dot {\bar c} + i\, \bar\beta\,e\,\gamma +\frac {1}{2} \bar\beta\,(p\cdot\psi) - 2 \bar\beta^2\,\beta   \,\chi
 -2\bar b\,\bar \beta\,\chi,
\end{eqnarray}
from the {\it two} expressions  for $Q_{(a)b}^{(1, 2)}$ that have been quoted in Eq. (13). We would like to lay emphasis 
on the fact that, in the proof of {\it absolute anticommutativity} of the nilpotent 
(anti-)BRST charges (i.e. $s_b Q_{ab}  = -\,i\, {\{Q_{ab}, Q_b}\} = 0,
s_{ab} Q_b  = -\,i\, {\{Q_b, Q_{ab}}\} = 0$), we have to utilize the beauty and strength of the CF-type restriction: $b + \bar b + 2\,\beta\bar\beta = 0$. As far as the nilpotency property is concerned, we discuss more about the conserved  (anti-) BRST charges and thier equivalent forms
in our Appendix A. In exactly similar fashion, we discuss a few relevant theoretical computations in Appendix B
about the property of absolute anticommutativity of the conserved and nilpotent charges that are related with the 
(anti-)BRST symmetry transformations (9) and (10).
 We shall capture all these features of the (anti-)BRST charges (for our {\it free} scalar and massless spinning
 relativistic particles) in the language of (anti-)chiral supervariable approach in the forthcoming Sec. 5.

\section {(Anti-)Chiral Supervariable Approach: Nilpotent (Anti-)BRST Symmetries for a Scalar Particle}

We exploit here the symmetry invariant (i.e. (anti-)BRST invariant) restrictions on the (anti-)chiral supervariables 
to derive the off-shell nilpotent (anti-)BRST symmetry transformations (2). In this connection, first of all, 
we generalize the basic and auxiliary variables of the starting Lagrangian (1) onto (1, 1)-dimensional {\it anti-chiral} supermanifold as
\begin{eqnarray}
&&x_\mu (\tau)\longrightarrow X_\mu (\tau, \bar\theta) = x_\mu (\tau) + \bar\theta\, R_\mu (\tau),\;\;
p_\mu (\tau)\longrightarrow P_\mu (\tau, \bar\theta) = p_\mu (\tau) + \bar\theta\, S_\mu (\tau),\nonumber\\
&&e(\tau)\;\;\longrightarrow  E(\tau, \bar\theta) = e(\tau) + \bar\theta\, f_1(\tau),\qquad
c(\tau)\;\;\longrightarrow  F(\tau, \bar\theta) = c(\tau) + i\,\bar\theta\, B_1(\tau),\nonumber\\
&&\bar c(\tau)\;\;\longrightarrow  \bar F(\tau, \bar\theta) = \bar c(\tau) + i\,\bar\theta\, B_2(\tau),\,\;
b(\tau)\;\;\longrightarrow  \tilde B(\tau, \bar\theta) = b(\tau) + \bar\theta\, f_2(\tau),
\end{eqnarray} 
where the (1, 1)-dimensional supermanifold is parametrized by
the superspace variables $(\tau, \bar\theta)$ and the set $(R_\mu (\tau), S_\mu (\tau), f_1 (\tau), f_2 (\tau))$
consists of fermionic {\it secondary} variables and the set $(B_1(\tau), B_2 (\tau))$ is made up of the bosonic {\it secondary} 
variables. All these secondary variables are function of $\tau $ and they are to be determined {\it precisely} in terms of the basic and 
auxiliary variables of the starting Lagrangian (1) by exploiting the theoretical strength of the (anti-)BRST (i.e. {\it quantum} gauge)
invariant restrictions on the supervariables (defined on the anti-chiral supermanifold).

One of the key ingredients of the (anti-)chiral superfield/supervariable approach is the requirement that 
{\it all} the {\it quantum} gauge (i.e. (anti-)BRST) invariant quantities must be independent of the ``soul'' coordinates 
$(\theta, \bar\theta)$ because these Grassmannian variables (i.e. $\theta, \bar\theta$) are {\it only} the mathematical artifacts which can {\it not}
be physically realized\footnote{In the older literature [28], the spacetime coordinates have been referred to as the ``body" coordinates 
and the Grassmannian variables have been christened as the ``soul" coordinates. The former can be realized {\it physically} 
but the {\it latter} variables are mathematical artifacts which can {\it not} be realized in the {\it same} way.}. 
We note that the following interesting quantities are BRST invariant, namely;
\begin{eqnarray} 
&&s_b b = 0,\qquad s_b p_\mu = 0, \qquad s_b c = 0,\qquad s_b (e\,\dot c) = 0,\nonumber\\
&&s_b (b\, e + i\,\bar c\,\dot c) = 0, \qquad\qquad s_b (\dot x_\mu - e\,p_\mu) = 0,
\end{eqnarray}
 where, in the last entry, we have to use the physical property of a {\it free} relativistic particle for which $\dot p_\mu = 0$.
 The above BRST-invariant quantities must be independent of $\bar\theta$ when {\it these} are generalized 
onto the (1, 1)-dimensional {\it anti-chiral} supermanifold; namely;
\begin{eqnarray} 
&&\tilde B(\tau, \bar\theta) = b(\tau), \quad P_\mu (\tau, \bar\theta) = p_\mu (\tau), \quad F(\tau, \bar\theta) = c(\tau),\nonumber\\
&& E(\tau, \bar\theta)\,\dot F(\tau, \bar\theta) = e(\tau)\,\dot c(\tau), \nonumber\\
&&B(\tau, \bar\theta)\, E(\tau, \bar\theta) + i\,\bar F(\tau, \bar\theta)\,\dot F(\tau, \bar\theta) 
= b(\tau)\, e(\tau) + i\,\bar c(\tau)\,\dot c(\tau), \nonumber\\
&&\dot X_\mu (\tau, \bar\theta) - E(\tau, \bar\theta)\,P_\mu (\tau, \bar\theta) = \dot x_\mu (\tau) - e(\tau)\,p_\mu (\tau). 
\end{eqnarray}
The above BRST-invariant restrictions yield the following expressions for the  secondary variables
 in terms of the basic and auxiliary  variables:
 \begin{eqnarray}
&& R_\mu(\tau) = c\, p_\mu,\qquad S_\mu (\tau) = 0,\qquad f_1(\tau) = \dot c,\nonumber\\
 &&B_1(\tau) =0,\qquad \quad B_2(\tau) = b,\qquad \quad f_2(\tau) = 0.
 \end{eqnarray}
 When we substitute these expressions into the super expansions (cf. Eq. (18)) of the anti-chiral supervariables, we obtain the following super expansions for the supervariables of our theory; namely;
 \begin{eqnarray}
&& P^{(b)}_\mu (\tau,\bar\theta) = p_\mu (\tau) + \bar \theta\, (0)\equiv p_\mu(\tau) +\bar\theta \,(s_b p_\mu),\nonumber\\
&&\bar F^{(b)}(\tau,\bar\theta) = \bar c +\bar\theta \,(ib)\equiv \bar c(\tau) + \bar\theta\, (s_b\bar c),\nonumber\\
&&F^{(b)}(\tau,\bar\theta) = c(\tau) + \bar\theta\,(0)\equiv c(\tau) + \bar\theta\,(s_b\, c),\nonumber\\
&&B^{(b)}(\tau,\bar\theta) = b(\tau) +\bar\theta\, (0)\equiv b(\tau) + \bar\theta \,(s_b b),\nonumber\\
&&E^{(b)}(\tau,\bar\theta) = e(\tau) + \bar\theta \,(\dot c)\equiv e(\tau) + \bar\theta \,(s_b e),\nonumber\\
&&X^{(b)}_\mu(\tau,\bar\theta) = x_\mu(\tau) + \bar\theta\,(c\,p_\mu)\equiv x_\mu (\tau) + \bar\theta \,(s_b x_\mu),
 \end{eqnarray}
 where the superscript $(b)$, on the supervariables, denotes the fact that these supervariables  have been 
 obtained after the application of  BRST invariant restrictions (20). It is evident that we have computed {\it all}
the BRST symmetry transformations $(s_b)$ for {\it all} the variables of the (anti-)BRST invariant Lagrangian (1) for a  {\it free}
 massive {\it scalar} 
relativistic particle. The super expansions in (22) $also$ establish a relationship  between  
$\partial_{\bar\theta}\equiv  {\partial}/{\partial\bar\theta}$ and the 
BRST symmetry transformation $s_b$  (e.g. $ \partial_{\bar\theta} \,X_\mu^{(b)} (\tau, \bar\theta) = s_b\, x_\mu,
\partial_{\bar\theta} \,\,\bar F^{(b)} (\tau, \bar\theta) = s_b \,\bar c,$ etc.) 
and, hence, the nilpotency $(\partial_{\bar\theta}^2 = 0, s_b^2 = 0)$
of these operators (in the superspace and ordinary space) are {\it also} inter-related .

To obtain the anti-BRST symmetry transformations that have been quoted in (2), we consider the generalizations of the 
basic and auxiliary variables of the starting Lagrangian (1) onto a (1, 1)-dimensional {\it chiral} supermanifold
(parameterized by the superspace coordinates $(\tau, \theta))$ 
as:
\begin{eqnarray}
&&x_\mu (\tau)\longrightarrow X_\mu (\tau, \theta) = x_\mu (\tau) + \theta\, \bar R_\mu (\tau),\nonumber\\
&&p_\mu (\tau)\longrightarrow P_\mu (\tau, \theta) = p_\mu (\tau) + \theta\,\bar S_\mu (\tau),\nonumber\\
&&e(\tau)\;\;\longrightarrow  E(\tau, \theta) = e(\tau) + \theta\,\bar f_1(\tau),\nonumber\\
&&c(\tau)\;\;\longrightarrow  F(\tau, \theta) = c(\tau) + i\,\theta\, \bar B_1(\tau),\nonumber\\
&&\bar c(\tau)\;\;\longrightarrow  \bar F(\tau, \theta) = \bar c(\tau) +i\, \theta\, \bar B_2(\tau),\nonumber\\
&&b(\tau)\;\;\longrightarrow  \tilde B(\tau, \bar\theta) = b(\tau) + \theta\,\bar f_2(\tau),
\end{eqnarray}
where the secondary variables $(\bar B_1(\tau), \bar B_2(\tau))$ are {\it bosonic} in nature and {\it fermi-onic} 
secondary variables of the above super expansions  are: $\bar R_\mu (\tau), \bar S_\mu(\tau), \bar f_1 (\tau), \bar f_2 (\tau)$.
These secondary variables are to be determined in terms of the basic and auxiliary variables of the starting Lagrangian 
(1) by exploiting the basic tenets of (anti-)chiral supervariable approach where we demand that the {\it quantum} gauge (i.e.
anti-BRST) invariant quantities should be independent of the ``soul" coordinate $\theta $. Towards this goal  in mind, we note that the 
following interesting quantities\footnote{We
would like to emphasize  that the (anti-)BRST invariant
quantities, listed in (24) and (19), have been obtained by the trial and error method
because there is {\it no} definite rule/principle to obtain them. More such kind 
of quantities {\it might} exist in the theory.} of the 1D toy model of a free {\it scalar} relativistic particle, namely;
\begin{eqnarray} 
&&s_{ab}\, \bar c = 0,\qquad s_{ab} \,b = 0, \qquad s_{ab}\, (e\,\dot{\bar c}) = 0,\qquad s_{ab} \,(b\, e + i\,\dot{\bar c}\,c) = 0,\nonumber\\
&&s_{ab} \,(\dot x_\mu - e\,p_\mu) = 0, \qquad\qquad s_{ab}\, p_\mu = 0,
\end{eqnarray}
are anti-BRST invariant. In particular, the last {\it but one} entry in the above equation is anti-BRST invariant because
we take into account the physical input for a 
{\it free} scalar relativistic particle where the force acting on the particle is zero (i.e. $\dot p_\mu = 0$)
which means that momentum is conserved.

We are now in the position to impose the following restrictions
on the supervariables   in accordance with the basic tenet of (anti-)chiral superfield approach to BRST formalism:
\begin{eqnarray} 
&&\bar F(\tau, \theta) = \bar c(\tau), \quad E(\tau, \theta)\,\dot 
{\bar F}(\tau, \theta) = e(\tau)\,\dot {\bar c}(\tau), P_\mu (\tau, \theta) = p_\mu (\tau) \nonumber\\
&&\tilde B(\tau, \theta)\, E(\tau, \theta) + i\,\dot {\bar F}(\tau, \theta) \,\, F(\tau, \theta)
= b(\tau)\, e(\tau) + i\,\dot {\bar c}(\tau)\,c(\tau),\nonumber\\
&&\dot X_\mu (\tau, \theta) - E(\tau, \theta)\,P_\mu (\tau, \theta) = \dot x_\mu (\tau) - e(\tau)\,p_\mu (\tau)\quad\tilde B(\tau, \theta) = b(\tau). 
\end{eqnarray}
The above restrictions lead to the derivation of the secondary variables in terms of the basic and auxiliary variables 
as:
\begin{eqnarray}
 &&\bar R_\mu(\tau) = \bar c\,\,\, p_\mu,\qquad\bar S_\mu (\tau) = 0,\qquad
\bar f_1(\tau) = 0,\nonumber\\
&&\bar B_2(\tau) = 0,\qquad\bar f_2(\tau) = \dot {\bar c},\qquad \bar B_1(\tau) =- b(\tau).
 \end{eqnarray}
 Thus, it is crystal clear that the secondary variables in the expansions (23) are found {\it accurately} in
 terms of the dynamical and auxiliary  variables of Lagrangian $L_b$ (cf. Eq. (1)).  
The substitution of  the above values into the super expansions (23) leads to the following
\begin{eqnarray}
 && X^{(ab)}_\mu(\tau,\theta) = x_\mu(\tau) + \theta\,(\bar c\,p_\mu)\equiv x_\mu (\tau) + \theta \,(s_{ab}\, x_\mu),\nonumber\\
 && P^{(ab)}_\mu (\tau,\theta) = p_\mu (\tau) +  \theta\, (0)\equiv p_\mu(\tau) +\theta \,(s_{ab}\, p_\mu),\nonumber\\
 &&F^{(ab)}(\tau,\theta) = c(\tau) + \theta\,(- i b)\equiv c(\tau) + \theta\,(s_{ab}\, c),\nonumber\\
 &&\bar  F^{(ab)}(\tau,\theta) = \bar c +\theta \,(0)\equiv \bar c(\tau) + \theta\, (s_{ab}\,\bar c),\nonumber\\
 &&E^{(ab)}(\tau,\theta) = e(\tau) + \theta \,(\dot{\bar c})\equiv e(\tau) + \theta \,(s_{ab}\, e),\nonumber\\
 &&\tilde B^{(ab)}(\tau,\theta) = b(\tau) + \theta\, (0)\equiv b(\tau) + \theta \,(s_{ab}\,b),
 \end{eqnarray}
where the superscript $(ab)$ on the supervariables denotes that these supervariables have been obtained after the application
of anti-BRST invariant restrictions (25).  
Moreover, we note that the coefficients of $\theta$, in the above super expansions, are nothing but the anti-BRST 
symmetry transformations (cf. Eq. (2)) for {\it all} the  basic and auxiliary variables of Lagrangian (1).
We note that: $\partial_\theta\longleftrightarrow s_{ab}$. In other words, the translational generator $\partial_\theta$
along $\theta$-direction of the {\it chiral} supermanifold is connected with the anti-BRST symmetry transformations
(i.e. $\partial_\theta F^{(ab)} (\tau, \theta) = s_{ab}\, c, \partial_\theta E^{(ab)} (\tau, \theta) = s_{ab}\, e$, etc.). Hence, 
the nilpotency property ($\partial_\theta^2 = 0, s_{ab}^2 = 0)$ of {\it both} these operators are inter-connected.
These nilpotency properties {\it also} imply the nilpotency of the conserved anti-BRST charges (cf. Sec. 5)
because $s_{(a)b}^2 = 0\,\Leftrightarrow Q_{(a)b}^2 = 0$.

\section{(Anti-)BRST Symmetries for a Spinning  Relativistic Particle: (Anti-)Chiral Supervariable Approach}

In this section, we exploit the idea of (anti-)BRST invariant restrictions on the specific combination(s) of the (anti-)chiral 
supervariables to derive the BRST and anti-BRST symmetry transformations for a 1D {\it free} massless
spinning relativistic particle. The basic concepts behind this 
theoretical trick is to demand that {\it all} the (anti-)BRST invariant quantities 
(which are physical quantities at the {\it quantum} level) should
be independent of the ``soul'' coordinates (i.e. Grassmannian variables)  $\theta$ and $\bar\theta$.
 Towards this objective in mind, first of all, we generalize 
{\it all} the variables of Lagrangian $ L_B $ (cf. Eq. (6)) on the appropriately chosen (1, 1)-dimensional 
{\it anti-chiral} supermanifold as follows:
\begin{eqnarray}
x_\mu (\tau)& \longrightarrow & X_\mu (\tau, \bar\theta) = x_\mu (\tau) + \bar\theta\, R_\mu (\tau),\nonumber\\
\gamma(\tau)\;\;& \longrightarrow & G (\tau, \bar\theta) = \gamma(\tau) + \bar\theta\, b_2 (\tau),\nonumber\\
p_\mu (\tau)& \longrightarrow & P_\mu (\tau, \bar\theta) = p_\mu (\tau) + \bar\theta\, S_\mu (\tau),\nonumber\\
c(\tau)\;\;\;& \longrightarrow & F(\tau, \bar\theta) = c(\tau) + i\,\bar\theta\, B_1(\tau),\nonumber\\
\psi_\mu (\tau) & \longrightarrow & \Psi_\mu (\tau, \bar\theta) = \psi_\mu (\tau) + \bar\theta\,B_\mu (\tau),\nonumber\\
\bar c(\tau)\;\; &\longrightarrow & \bar F(\tau, \bar\theta) = \bar c(\tau) + i\,\bar\theta\, B_2(\tau),\nonumber\\
e(\tau)\;\;& \longrightarrow & E(\tau, \bar\theta) = e(\tau) + \bar\theta\, f_1(\tau),\nonumber\\
\beta (\tau)\;\;\, & \longrightarrow &\tilde\beta (\tau, \bar\theta) = \beta(\tau) + i\,\bar\theta\, f_3(\tau),\nonumber\\
b(\tau)\;\;& \longrightarrow &  \tilde B(\tau, \bar\theta) = b(\tau) + \bar\theta\, f_2(\tau),\nonumber\\
\bar\beta (\tau)\;\;& \longrightarrow & \tilde{\bar\beta} (\tau, \bar\theta) =\bar \beta(\tau) + i\,\bar\theta\, f_4(\tau),\nonumber\\
\chi(\tau)\;\;& \longrightarrow & X (\tau, \bar\theta) = \chi(\tau) + \bar\theta\, b_1 (\tau ),\nonumber\\
 \bar b(\tau)\;\;& \longrightarrow &\tilde {\bar B}
(\tau, \bar\theta) = \bar b (\tau) + \bar\theta f_5 (\tau), 
\end{eqnarray} 
where the superspace coordinates  $(\tau, \bar\theta)$ characterize the (1, 1)-dimensional anti-chiral supermanifold
and secondary variables ($R_\mu, S_\mu, f_1, f_2, f_3, f_4, f_5$) are fermionic  and the set ($B_\mu, b_1, b_2, B_1, B_2$)
is bosonic in nature. The secondary variables are to be determined   in terms of the basic and auxiliary variables
of the Lagrangian $L_B$ by exploiting the BRST invariant restrictions for the derivation of BRST symmetry transformations (10).

The basic tenet of (anti-)chiral supervariable approach demands that the BRST invariant quantities 
should be independent of the Grassmannian variable $\bar\theta$ (of the anti-chiral supermanifold)
when  they are generalized onto the (1, 1)-dimensional anti-chiral supermanifold. It is straightforward 
to note that the {\it trivially} invariant quantities such as: $s_b b = 0, s_b\,\gamma  = 0, s_b\,\beta  = 0,
s_b \;p_\mu = 0$ imply that  $B(\tau, \bar\theta) = b(\tau), G(\tau, \bar\theta) = \gamma (\tau),
\tilde \beta (\tau, \bar\theta) = \beta (\tau), P_\mu (\tau, \bar\theta) = p_\mu (\tau)$.   
The above restrictions lead to the following trivial expressions for the  secondary variables in the above
anti-chiral super expansions of the supervariables (cf. Eq. (28)):
\begin{eqnarray}
f_2 (\tau) = 0,\qquad b_2(\tau) = 0,\qquad f_3 (\tau) = 0,\qquad S_\mu (\tau) = 0.
\end{eqnarray}
In other words, we have the following super expansions 
\begin{eqnarray}
&& B^{(b)}(\tau, \bar\theta) = b(\tau) + \bar\theta\,(0) = b(\tau) + \bar\theta \;(s_b b),\nonumber\\
&& G^{(b)}(\tau, \bar\theta) = \gamma (\tau) + \bar\theta\, (0) = 
\gamma(\tau) + \bar\theta\; (s_b \gamma),\nonumber\\
&&\tilde \beta^{(b)}(\tau, \bar\theta) =  \beta (\tau) +  \bar\theta\, (0) =  \beta (\tau) +  \bar\theta\; (s_b\beta),\nonumber\\
&&P_\mu^{(b)}(\tau, \bar\theta) =  p(\tau) +  \bar\theta\, (0) =  \beta (\tau) +  \bar\theta\; (s_b\,p_\mu),
\end{eqnarray}
where the superscript $(b)$ denotes the expansions for the supervariables  after the application of 
BRST invariant restrictions and the coefficients of $\bar\theta$, in the above expansions,  
are nothing but the BRST symmetry transformations (i.e. $s_b b = 0, s_b \gamma = 0, s_b\,\beta = 0, s_b\,p_\mu = 0$)
that have been quoted in Eq. (10). In other words, we have already derived the BRST symmetry transformations
for the variables ($b, \gamma,\beta, p_\mu $) of our theory of a massless {\it spinning} particle.

We now concentrate on the derivation of the BRST symmetry transformations
for some  non-trivial variables of our 1D toy model of a free {\it spinning}  relativistic particle.
Towards this goal in mind, we note that the following BRST-invariant quantities\footnote{It should be noted 
that $s_b (\dot x_\mu -\,e\,p_\mu + i\,\chi\,\psi_\mu = 0$) is valid {\it only}   when we take the on-shell
conditions: $\dot p_\mu = 0, \dot \psi_\mu = \chi\,p_\mu$ where $\dot p_\mu = 0$ implies the {\it free} motion of 
the particle where the force acting on it is {\it zero} and $\dot {\psi_\mu} = \chi\,p_\mu$ 
implies $p\cdot\dot\psi = \chi \;p^2\approx 0$ which  {\it basically }
provides the connection between  the constraints ($ p^2 \approx 0, \; p \cdot \psi \approx 0$)
on our system of a {\it free  massless} spinning particle where we have {\it also} taken $\dot p_\mu = 0$.}. 
\begin{eqnarray}
&&s_b(\bar\beta\;\gamma) = 0,\qquad\qquad s_b (b\,\bar\beta + \gamma\,\bar c) = 0, \qquad\qquad s_b(\beta^2 \bar\beta  + c\,\gamma) = 0,\qquad
\nonumber\\
&&s_b(\dot c + 2\beta\chi) = 0,\qquad s_b(\bar b + 2\beta\bar\beta) = 0,\qquad s_b( e\,\gamma\,\chi + e\,\bar\beta\dot\beta - i\;\bar\beta\dot c \chi) = 0,\nonumber\\
&&s_b(\dot x_\mu  - e\,p_\mu  + i\,\chi\,\psi_\mu) = 0,\qquad\qquad s_b(c\,p_\mu + \beta\psi_\mu) = 0,
\end{eqnarray}
are very useful for us because these can be generalized onto (1, 1)-dimensional anti-chiral supermanifold and we can invoke, for the 
following {\it simple} case, the basic tenet of 
(anti-) chiral supervariable approach to BRST formalism and demand the following
\begin{eqnarray}
\tilde{\bar\beta}(\tau, \bar\theta)\;G^{(b)}(\tau, \bar\theta) = \bar\beta(\tau)\,\gamma (\tau),\qquad\qquad G^{(b)} (\tau, \bar\theta) = \gamma (\tau),
\end{eqnarray}
which implies that $f_4 (\tau)\propto \gamma$. We choose, for the sake of brevity: $f_4 (\tau) =  \gamma (\tau)$.
This yields:
\begin{eqnarray}
\bar\beta ^{(b)} (\tau, \bar\theta) = \bar\beta (\tau) + \bar\theta\;(i\,\gamma)\equiv \bar\beta (\tau) + \bar\theta\,(s_b\, \bar\beta). 
\end{eqnarray}  
It is clear that the coefficient of $\bar\theta$ is the BRST symmetry transformation on $\bar\beta$ (cf. Eq. (10)).
The above equation (33) would be used in the following equality due to our observation in Eq. (31) that 
$s_b (b\,\bar \beta + \gamma\,\bar c) = 0$, namely;
\begin{eqnarray}
\tilde B^{(b)} (\tau, \bar\theta) \bar\beta ^{(b)}(\tau, \bar\theta) + G^{(b)}(\tau, \bar\theta)\bar F (\tau, \bar\theta) = 
b(\tau)\bar\beta(\tau) + \gamma(\tau)\,\bar c(\tau),
\end{eqnarray}
which leads to the derivation of $B_2 (\tau) = b(\tau)$
where we have used the expansion  of $\bar F (\tau, \bar\theta)$ from (28) and taken the inputs: $G^{(b)}(\tau, \bar\theta) = \gamma (\tau),\,
\tilde B^{(b)}(\tau, \bar\theta) = b(\tau)$ (cf. Eq. (30)). Finally, we obtain the following super expansion for the supervariable $\bar F (\tau, \bar\theta)$:
\begin{eqnarray}
\bar F^{(b)}(\tau, \bar\theta) = \bar c(\tau) + \bar\theta\;(i\,b(\tau))\equiv \bar c (\tau)  + \bar\theta\,(s_b\, \bar c).
\end{eqnarray}
The above expansion leads to $s_b \bar c = i\,b$ (cf. Eq. (10)) as the BRST symmetry transformation
 on the variable $\bar c(\tau)$. Similar sets of exercises lead to the derivation of 
secondary variables of the super expansions (28) in terms the basic and auxiliary variables of BRST invaraint Lagrangian $L_B$ (for the free 
motion of a massless {\it spinning} relativistic particle) as:
\begin{eqnarray}
&&B_1  = -\,\beta^2,\qquad B_\mu  = i\,\beta\,p_\mu, \qquad b_1  = i\,\dot\beta,\qquad f_5 = - 2 \,i\,\beta\,\gamma,
 \nonumber\\
&&f_1 = \dot c + 2\beta\chi,\quad\qquad R_\mu  = c\;p_\mu + \beta\;\psi_\mu.
\end{eqnarray}
We would like to lay emphasis on the fact that we have freely used the force-free condition ($\dot p_\mu = 0$) and the on-shell
conditions (e.g. $\dot \psi_\mu = \chi \, p_\mu$) in the derivation of the last entry in the above equation.
The substitution of {\it all} these values of the secondary variables (in terms of the basic auxiliary variables
of the Lagrangian $L_B$) into the super expansions of {\it all}
the supervariables (28) leads to
the following  expansions (with BRST symmetry transformations $s_b$ as given in Eq. (10)):
\begin{eqnarray}
&&X_\mu^{(b)} (\tau, \bar\theta) = x_\mu (\tau) + \bar\theta\, (c\,p_\mu + \beta\,\psi_\mu)\equiv x_\mu(\tau) + \bar\theta\,(s_b \,x_\mu),\nonumber\\
&&P_\mu^{(b)} (\tau, \bar\theta) = p_\mu (\tau) + \bar\theta\,(0) \equiv p_\mu (\tau)  + \bar\theta\,(s_b\, p_\mu),\nonumber\\
&&\Psi_\mu^{(b)} (\tau, \bar\theta) = \psi_\mu (\tau) + \bar\theta\,(i\,\beta\,p_\mu)\equiv \psi_\mu (\tau) + \bar\theta\, (s_b\, \psi_\mu),\nonumber\\
&&F ^{(b)}(\tau, \bar\theta) = c(\tau) + \bar\theta\, (-\,i\,\beta^2)\equiv c(\tau) + \bar\theta\,(s_b\, c),\nonumber\\
&&\bar F^{(b)}(\tau, \bar\theta) = \bar c(\tau) + \bar\theta\,(i\,b)\equiv \bar c(\tau) + \bar\theta\,(s_b\,\bar c),\nonumber\\
&&\tilde B^{(b)}(\tau, \bar\theta) = b(\tau) + \bar\theta\,(0)\equiv b(\tau) + \bar\theta\,(s_b\, b),\nonumber\\
&&\tilde {\bar B}^{(b)}(\tau, \bar\theta) = \bar b (\tau) + \bar\theta (-2\,i\,\beta\,\gamma)\equiv \bar b(\tau) + \bar\theta\,(s_b \,\bar b),\nonumber\\
&&G^{(b)} (\tau, \bar\theta) = \gamma(\tau) + \bar\theta\,(0)\equiv \gamma (\tau) +\bar\theta\, (s_b\, \gamma),\nonumber\\
&&E^{(b)}(\tau, \bar\theta) = e(\tau) + \bar\theta\,(\dot c + 2\beta\,\chi)\equiv e(\tau) + \bar\theta\,(s_b\, e),\nonumber\\
&&\tilde\beta^{(b)} (\tau, \bar\theta) = \beta(\tau) + \bar\theta\,(0)\equiv \beta (\tau) + \bar\theta\, (s_b\,\beta),\nonumber\\
&& \tilde{\bar\beta}^{(b)} (\tau, \bar\theta) =\bar \beta(\tau) + \bar\theta\,(i\,\gamma)\equiv \bar\beta(\tau) + \bar\theta\, (s_b\, \bar\beta),\nonumber\\
&&X^{(b)} (\tau, \bar\theta) = \chi(\tau) + \bar\theta\,(i\,\dot\beta)\equiv \chi (\tau) + \bar\theta \,(s_b\, \chi).
 \end{eqnarray} 
We note that we have derived {\it all} the BRST symmetry transformations (cf. Eq. (10)) as the coefficients of $\bar\theta$ in the above {\it total} super expansions for our present 1D theory when  it is generalized onto (1, 1)-dimensional anti-chiral supermanifold. A close look
at (37) shows that we have a relationship\footnote{We have taken the {\it partial} derivative $\partial_{\bar\theta}$ because the 
(1, 1)-dimensional (anti-)chiral supermanifold is parameterized by superspace coordinates $(\tau, \bar \theta)$ 
which are two in numbers.}: $\partial_{\bar\theta}\longleftrightarrow s_b $. In other words, there is a deep connection   
between the  BRST symmetry transformation $s_b$ in the ordinary 1D space {\it and} the geometrical quantity (i.e. translational generator $\partial_{\bar\theta}$)
on the (1, 1)-dimensional {\it anti-chiral} supermanifold (parameterized by $(\tau, \bar\theta$)).

We can concentrate now on the derivation of the nilpotent anti-BRST symmetry transformations 
(cf. Eq. (9)) by exploiting the beauty and strength of the (anti-)chiral supervariable formalism
where we demand that {\it all} the anti-BRST invariant quantities (i.e. all the ``physical"
quantities at the {\it quantum} level) must be independent of the ``soul'' coordinate $\theta$ of the 
(1, 1)-dimensional {\it chiral} supermanifold on which {\it all} the variables of the Lagrangian $L_{\bar B}$
(cf. Eq. (6)) are generalized as:
\begin{eqnarray}
x_\mu (\tau)& \longrightarrow & X_\mu (\tau, \theta) = x_\mu (\tau) + \theta\, \bar R_\mu (\tau),\nonumber\\
\gamma(\tau)\;\;& \longrightarrow & G (\tau, \theta) = \gamma(\tau) + \theta\, \bar b_2 (\tau),\nonumber\\
p_\mu (\tau)& \longrightarrow & P_\mu (\tau, \theta) = p_\mu (\tau) + \theta\, \bar S_\mu (\tau),\nonumber\\
c(\tau)\;\;\; & \longrightarrow & F(\tau, \theta) = c(\tau) + i\,\theta\, \bar B_1(\tau),\nonumber\\
\psi_\mu (\tau)& \longrightarrow &  \Psi_\mu (\tau, \theta) = \psi_\mu (\tau) + \theta\,\bar B_\mu (\tau),\nonumber\\
\bar c(\tau)\;\;& \longrightarrow &   \bar F(\tau, \theta) = \bar c(\tau) + i\,\theta\,\bar B_2(\tau),\nonumber\\
e(\tau)\;\; & \longrightarrow & E(\tau, \theta) = e(\tau) + \theta\, \bar f_1(\tau),\nonumber\\
 \beta (\tau)
 & \longrightarrow & \tilde\beta (\tau, \
\theta) = \beta(\tau) + i\,\theta\, \bar f_3 (\tau),\nonumber\\
b(\tau)\;\; & \longrightarrow &  \tilde B(\tau, \theta) = b(\tau) + \theta\,\bar f_2(\tau),\nonumber\\
\bar\beta (\tau)\;\; & \longrightarrow & \tilde{\bar
\beta} (\tau, \theta) =\bar \beta(\tau) + i\,\theta\, \bar f_4(\tau),\nonumber\\
\chi(\tau)\;\; & \longrightarrow & X (\tau, \theta) = \chi(\tau) + \theta\,\bar b_1 (\tau ),\nonumber\\
 \bar b(\tau)\;\; & \longrightarrow & \tilde {\bar B}
(\tau, \theta) = \bar b (\tau) + \theta \bar f_5 (\tau), 
\end{eqnarray}
where the present (1, 1)-dimensional   {\it chiral} supermanifold is 
parameterized by the superspace variable $(\tau, \theta)$ and the secondary variables, on the r.h.s. of the 
above super expansions, are fermionic (i.e. $\bar R_\mu (\tau), \bar S_\mu(\tau), \bar f_1 (\tau), \bar f_2 (\tau),\bar f_3 (\tau), \bar f_4(\tau),
\bar f_5 (\tau)$) as well as bosonic $(\bar B_\mu (\tau), \bar b_1 (\tau), \bar b_2 (\tau), \bar B_1 (\tau), \bar B_2(\tau))$ in nature.
These secondary variables would be determined by exploiting the idea of anti-BRST invariant restrictions on the
supervariables.

One of the fundamental concepts behind supervariable/superfield approach to BRST formalism is the requirement that
{\it all} the (anti-)BRST invariant quantities should be independent of the Grassmannian variables when they are generalized 
onto  appropriately chosen (anti-)chiral supermanifolds.
It is straightforward to note that the {\it trivial} anti-BRST invariant quantities such as: 
$s_{ab}\bar b = 0,\, s_{ab} \gamma = 0, s_{ab}\,\bar\beta  = 0,\, s_{ab}\,p_\mu = 0$ imply  that $\tilde {\bar B}(\tau, \theta) = \bar b(\tau), 
G(\tau, \theta) = \gamma(\tau), \tilde{\bar\beta}(\tau, \theta) = \bar\beta(\tau), P_\mu (\tau, \theta) = p_\mu(\tau)$.
These restrictions, in turn, imply that the secondary variables of the above {\it chiral} supervariables, in the expansions (38), are zero.
Thus, we have the following {\it trivial} values of the secondary variables:
\begin{eqnarray}
\bar f_5(\tau) = 0, \qquad \bar b_2(\tau) = 0,\qquad \bar f_4(\tau) = 0, \qquad \bar S_\mu(\tau)  = 0.
\end{eqnarray}
As a consequence of the above, we have the following {\it chiral} super expansions  
\begin{eqnarray}
&&\tilde{\bar B}^{(ab)}(\tau, \theta)  = \bar b(\tau) + \theta \;(0) \equiv \bar b(\tau) + \theta \;(s_{ab} \bar b),\nonumber\\
&& G^{(ab)}(\tau, \theta)  = \gamma(\tau) + \theta \;(0) \equiv \gamma(\tau) + \theta \;(s_{ab} \gamma),\nonumber\\
&&\tilde{\bar \beta}^{(ab)}(\tau, \theta)  = \bar \beta(\tau) + \theta \;(0) \equiv \bar \beta(\tau) + \theta \;(s_{ab} \bar\beta),\nonumber\\
&&P_\mu^{(ab)}(\tau, \theta)  = p_\mu(\tau) + \theta \;(0) \equiv p_\mu(\tau) + \theta \;(s_{ab} p_\mu),
\end{eqnarray}
where the superscript $(ab)$ denotes the {\it chiral} super expansions of the chiral supervariables  after the application of the anti-BRST
invariant restrictions. We observe that the coefficients of $\theta$, in the above expansions, are nothing but the 
anti-BRST symmetry transformations (9) for the variables ($\bar b, \gamma, \bar\beta, p_\mu$) of our theory which is described
 by the Lagrangian $L_{\bar B}$ (cf. Eq. (6)). We shall exploit the expansions (40) for our further discussions.

To utilize the potential and power of the (anti-)chiral supervariable approach to BRST formalism,
it is  very important for us to obtain the useful (anti-)BRST invariant quantities because it is primarily 
these quantities that are generalized onto appropriately chosen  supermanifold where we demand 
the (anti-)BRST invariant (i.e. {\it quantum} gauge invariant) restrictions on the supervariables which lead to the 
derivation of  secondary variables in terms of the basic and auxiliary variables of the 
specifically chosen Lagrangian of the theory. In this context, we note that the following useful and interesting
anti-BRST invariant quantities 
\begin{eqnarray}
&&s_{ab}\,(\beta\;\gamma) = 0,\quad\quad s_{ab}\, (\bar b\,\beta - \gamma\, c) = 0, \quad\quad 
s_{ab}\,(\beta \bar\beta^2  - \bar c\,\gamma) = 0,\qquad
\nonumber\\
&&s_{ab}\,(\dot {\bar c} + 2\,\bar\beta\chi) = 0,\; s_{ab}\,( b + 2\beta\bar\beta) = 0,\; s_{ab}\,(  e\,\gamma\,\chi - e\,
\dot {\bar\beta}\,\beta + i\;
\beta\,\dot {\bar c}\, \chi) = 0,\nonumber\\
&&s_{ab}\,(\dot x_\mu  - e\,p_\mu  + i\,\chi\,\psi_\mu) = 0,\qquad\qquad s_{ab}\,(\bar c\,p_\mu + \bar\beta\psi_\mu) = 0,
\end{eqnarray}    
 are of paramount importance to us because these can be generalized onto the (1, 1)-dimensional {\it chiral}
 supermanifold in terms of the chiral supervariables which would be subjected to the restrictions that these  quantities {\it must} be 
 {\it independent} of the Grassmannian variable $\theta$. To elaborate on it, we start off with a simple supervariable restriction
 where we demand that
 \begin{eqnarray}
 \tilde\beta (\tau,\theta) \,G^{(ab)}(\tau,\theta) = \beta(\tau)\,\gamma(\tau),
 \end{eqnarray}
where $\tilde\beta (\tau,\theta)$ and $G^{(ab)}(\tau,\theta) $ are given in 
expansions (38) and (40), respectively, which have to be utilized in our equation (42). We observe that the above equality
leads to: $\bar f_3(\tau)\propto \gamma$. We choose for the sake of brevity: $\bar f_3 (\tau) = - \gamma$ (which is
different from our earlier choice $f_4(\tau) = + \gamma$ in the context of the derivation of {\it BRST symmetry}
transformation for $\bar\beta(\tau)$).
This choice immediately  implies that we have derived the following:
\begin{eqnarray}
\tilde \beta^{(ab)}(\tau,\theta)  = \beta (\tau) + \theta \,(- \,i\,\gamma)\equiv \beta (\tau) + \theta\, (s_{ab}\,\beta).
\end{eqnarray}  
The above super expansion states clearly that we have {\it already} derived the anti-BRST symmetry transformation
($s_{ab}\;\beta = - i\;\gamma$) as the coefficient of $\theta$ (cf. Eq. (9)). 
We utilize the above expansions (cf. Eqs. (40), (43)) in the anti-BRST invariant quantity [$s_{ab}(\bar b\, \beta - \gamma\, c) = 0$]
where we also take  the help of $\tilde{\bar B}^{(ab)}(\tau,\theta) = \bar b (\tau) + \theta \;(0)$ from Eq. (40). Finally, we have the 
following equality (which is nothing but an anti-BRST invariant restriction):
\begin{eqnarray}
 \tilde{\bar B}^{(ab)}(\tau,\theta)\tilde\beta^{(ab)}(\tau,\theta) - G^{(ab)}(\tau,\theta) F(\tau,\theta) = \bar b(\tau) \beta(\tau) - \gamma(\tau)
\, c(\tau).
\end{eqnarray} 
We know that $\tilde{\bar B}^{(ab)} (\tau,\theta) = \bar b(\tau), G^{(ab)}(\tau,\theta) = \gamma (\tau)$ and
the expansion for $\tilde\beta^{(ab)}(\tau,\theta)$ is given in Eq. (43). Using these inputs and super expansion
for $ F(\tau,\theta)$ from Eq. (38), it is evident that $\bar B_1 (\tau) = \bar b$.
Thus, we have obtained the following super expansion, namely;
\begin{eqnarray}
 F^{(ab)}(\tau,\theta)  =  c (\tau)  + \theta\,(i\, \bar b)\equiv  c(\tau)  + \theta\,(s_{ab}\, c(\tau)),
\end{eqnarray}
where the superscript $(ab)$ denotes the super expansion of the 
supervariable $ F (\tau, \theta)$ after the application of anti-BRST invariant restrictions 
(cf. Eq. (41) for details).

We have derived the anti-BRST symmetry transformations for the variables $\beta(\tau)$ and $ c(\tau)$ 
in the above equations (43) and (45), respectively. This exercise can be repeated with {\it all} the other anti-BRST invariant restrictions
that have been listed in equation (41). Ultimately, we obtain the following expressions  for the rest of the secondary 
variables in terms of the basic and auxiliary variables of $L_{\bar B}$ (cf. Eq. (6)), namely;
\begin{eqnarray}
&&\bar R_\mu = \bar c\,p_\mu + \bar\beta\psi_\mu,\qquad \bar B_\mu  = i\,\bar\beta\, p_\mu\qquad \bar f_1  = \dot{\bar c} + 2\,\bar\beta\chi,\nonumber\\
&&\bar b_1 = i\,\dot{\bar \beta},\qquad \bar B_2  = -\; \bar \beta^2,\qquad \bar f_5 = 2\,i\,\bar\beta\,\gamma.
\end{eqnarray}   
Substitution of the secondary variables into the super expansions (38) and taking the help of super expansions (40), (43) and (45), we 
write the {\it final} super expansions for {\it all} the supervariables of our theory as  
\begin{eqnarray}
&&X_\mu^{(ab)} (\tau, \theta) = x_\mu (\tau) + \theta\, (\bar c\,p_\mu + \bar\beta\,\psi_\mu)\equiv x_\mu(\tau) + \theta\,(s_{ab} \,x_\mu),\nonumber\\
&&P_\mu^{(ab)} (\tau, \theta) = p_\mu (\tau) + \theta\,(0) \equiv p_\mu (\tau)  + \theta\,(s_{ab}\, p_\mu),\nonumber\\
&&\Psi_\mu^{(ab)} (\tau, \theta) = \psi_\mu (\tau) + \theta\,(i\,\bar\beta\,p_\mu)\equiv \psi_\mu (\tau) + \theta\, (s_{ab}\, \psi_\mu),\nonumber\\
&&F ^{(ab)}(\tau, \theta) = c(\tau) + \theta\, (i\,\bar b)\equiv c(\tau) + \theta\,(s_{ab}\, c),\nonumber\\
&&\bar F^{(ab)}(\tau, \theta) = \bar c(\tau) + \theta\,(-i\,\bar\beta^2)\equiv \bar c(\tau) + \theta\,(s_{ab}\,\bar c),\nonumber\\
&&\tilde B^{(ab)}(\tau, \theta) = b(\tau) + \theta\,(2\,i\,\bar\beta\,\gamma)\equiv b(\tau) + \theta\,(s_{ab}\, b),\nonumber\\
&&\tilde {\bar B}^{(ab)}(\tau, \theta) = \bar b (\tau) + \theta (0)\equiv \bar b(\tau) + \theta\,(s_{ab} \,\bar b),\nonumber\\
&&G^{(ab)} (\tau, \theta) = \gamma(\tau) + \theta\,(0)\equiv \gamma (\tau) +\theta\, (s_{ab}\, \gamma),\nonumber\\
&&E^{(ab)}(\tau, \theta) = e(\tau) + \theta\,(\dot {\bar c } + 2\bar\beta\,\chi)\equiv e(\tau) + \theta\,(s_{ab}\, e),\nonumber\\
&&\tilde\beta^{(ab)} (\tau, \theta) = \beta(\tau) + \theta\,(-\,i\,\gamma)\equiv \beta (\tau) + \theta\, (s_{ab}\,\beta),\nonumber\\
&& \tilde{\bar\beta}^{(ab)} (\tau, \theta) =\bar \beta(\tau) + i\,\theta\,(0)\equiv \bar\beta(\tau) + \theta\, (s_{ab}\, \bar\beta),\nonumber\\
&&X^{(ab)} (\tau, \theta) = \chi(\tau) + \theta\,(i\,\dot{\bar\beta})\equiv \chi (\tau) + \theta \,(s_{ab}\, \chi),
 \end{eqnarray} 
where the superscript $(ab)$ on supervariables denotes the super expansions of the supervariables after the application of the
anti-BRST invariant  restrictions (41). It is now {\it obvious} that we have derived {\it all} the anti-BRST symmetry transformation of Eq. (9)
as the coefficients of $\theta$ in the expansions (47). We note that we have also obtained a mapping $s_{ab}\longleftrightarrow \partial_\theta$
which states that the anti-BRST symmetry transformation ($s_{ab}$) in the {\it ordinary} 1D space is deeply connected with the 
translational generator ($\partial_\theta$) on the (1, 1)-dimensional {\it chiral} supermanifold.
Hence, their nilpotency properties  ($s_{ab}^2 = 0, \partial_\theta^{2} = 0$) in the superspace (of the chiral supermanifold)
and ordinary space (of the 1D flat spacetime manifold)
are also inter-related.

\section{(Anti-)BRST Charges: Off-Shell Nilpotency and Absolute Anticommutativity Properties}

In this section, we take up the {\it cases} of {\it both}: a {\it free} massive scalar as well as a {\it massless spinning} relativistic
particles and capture the nilpotency and absolute anticommutativity of their (anti-)BRST charges within the framework 
of (anti-)chiral supervariable approach to BRST formalism. First of all, we take up the case of a massive free {\it scalar} relativistic particle 
and focus on the nilpotent (anti-) BRST charges that have been expressed in Eq. (4). In particular, we concentrate on:
$Q_b = b\,\dot c- \dot b\,c$ and $Q_{ab} = b\;\dot {\bar c} -\dot b\,\bar c$. It can be checked that
these charges can be expressed in terms of  the (anti-)chiral supervariables and other mathematical quantities 
(that are defined on the (1, 1)-dimensional (anti-)chiral supermanifolds) as:
\begin{eqnarray}
Q_b  & = &\frac {\partial}{\partial\bar\theta}\;\Big[i\;\dot{\bar F}^{(b)}(\tau,\bar\theta)\, F^{(b)}(\tau,\bar\theta)
- i\; \bar F^{(b)}(\tau,\bar\theta)\,\dot F^{(b)}(\tau,\bar\theta)\Big]\nonumber\\
&\equiv & \int\; d \bar\theta\;\;\Big[i\;\dot{\bar F}^{(b)}(\tau,\bar\theta)\, F^{(b)}(\tau,\bar\theta)
- i\; \bar F^{(b)}(\tau,\bar\theta)\,\dot F^{(b)}(\tau,\bar\theta)\Big],
\end{eqnarray} 
\begin{eqnarray}
 Q_{ab}  & = & \frac {\partial}{\partial\theta}\;\Big[ i\;\bar F^{(ab)}(\tau,\theta)\;\dot F^{(ab)}(\tau,\theta)  
- i\;\dot {\bar F}^{(ab)}(\tau,\theta)\;F^{(ab)}(\tau,\theta)\Big]\nonumber\\
&\equiv & \int\; d\theta\;\;\Big[ i\;\bar F^{(ab)}(\tau,\theta)\;\dot F^{(ab)}(\tau,\theta)  
- i\;\dot {\bar F}^{(ab)}(\tau,\theta)\;F^{(ab)}(\tau,\theta)\Big],
\end{eqnarray}
where we have utilized the derivatives $(\partial_{\theta}, \partial_{\bar\theta}$) as well as differentials 
($d\theta, d\bar\theta$)  along with the supervariables (22) and (27) that have been derived after the application of the 
BRST and anti-BRST invariant restrictions (cf. Eqs. (20) and (25)) on the (anti-)chiral supervariables.
It is evident that we have the following {\it trivial} equalities:
\begin {eqnarray}
&& \partial_{\bar\theta}\;Q_b = 0,  \qquad \quad     \partial_{\theta}\;Q_{ab} = 0\qquad\Longleftrightarrow\qquad \partial_{\bar\theta}^2 =\partial_{\theta}^2 = 0.
\end{eqnarray} 
In other words, we note that the application of the derivatives w.r.t. Grassmannian variables $(\bar\theta)\theta$ 
(which characterize the (anti-)chiral supermanifolds) on the fermionic (i.e. off-shell nilpotent)
BRST and anti-BRST charges leads to {\it zero} result
due to the nilpotency of the translational generators $(\partial_{\bar\theta})\partial_{\bar\theta}$
along the $(\bar\theta)\theta$-directions of the (anti-)chiral supermanifolds.

The consequences of our observations in   (48), (49) and (50) become very transparent when we express {\it these} 
in the {\it ordinary} space with the help of the mappings: $s_b\longleftrightarrow \partial_{\bar\theta}$
 and $s_{ab}\longleftrightarrow \partial_\theta $. In other words, we observe the following interesting relationships,
 namely;
\begin{eqnarray}
&&Q_b = s_b \;[i\,\dot{\bar c}\,c - i\,\bar c\,\dot c] \;\;\;\;\;\qquad\qquad\quad \Longleftrightarrow \qquad s_b Q_b  = 0\;\; \;\;\Longleftrightarrow
\;\;\;s_b^2 = 0,\nonumber\\
&&Q_{ab} = s_{ab} \;[i\,\bar c\,\dot c - i\,\dot{\bar c}\, c]\;\;\;\;\;\;\;\;\;\quad\qquad \Longleftrightarrow\qquad  s_{ab} \,Q_{ab} = 0
\Longleftrightarrow \;\;\; s_{ab}^2 = 0,\nonumber\\
&&s_b Q_b= 0 \Longrightarrow  -i\,{\{Q_b,Q_b}\} = 0 \quad \;\;\;\;\Longrightarrow  \;\;\;\qquad\; Q_b^2 = 0 \;\;\;\Longleftrightarrow \;\;\;s_b^2 = 0,\nonumber\\
&&s_{ab} \,Q_{ab}= 0 \Longrightarrow -i\,\{Q_{ab},Q_{ab}\} = 0 \; \Longrightarrow \;\;\qquad Q_{ab}^2 = 0
\quad\Longleftrightarrow \;\;\;s_{ab}^2 = 0, 
\end{eqnarray}
which demonstrate that the nilpotency of the (anti-)BRST symmetries (i.e. $s_{(a)b}^2 = 0)$ implies the 
nilpotency of the (anti-)BRST charges (i.e. $Q_{(a)b}^2 = 0$). In other words, we note that the nilpotency
of the translational generators $\partial_{\bar\theta}^2 = 0, \partial_\theta^2 = 0$ (within the framework of 
(anti-)chiral supervariable approach) is intimately connected with the nilpotency of the (anti-)BRST symmetries and corresponding charges
in the {\it ordinary} space. This statement is corroborated by our observations in (48), (49), (50) and (51)
where the (anti-) BRST symmetries  and corresponding charges in the {\it ordinary} space are connected with the translational 
generators ($\partial_\theta, \partial_{\bar\theta}$) on the (1, 1)-dimensional (anti-)chiral supermanifolds.

We dwell a bit on the property of absolute anticommutativity of the off-shell nilpotent (anti-)BRST charges which is 
{\it one } of the key and {\it novel } observations of our present investigation. It can be checked, 
in the context of expansions in Eqs. (22) and (27), that the conserved (anti-)BRST charges can be {\it also} 
expressed as
\begin{eqnarray}
Q_{ab}  &= &\frac {\partial}{\partial\bar\theta}\;\Bigl [-i \; \bar F^{(b)}(\tau,\bar \theta)\; \dot {\bar F}^{(b)}(\tau, \bar\theta)
\Big] \equiv  \int\; d\bar\theta\;\Bigl [- i\; \bar F^{(b)}(\tau, \bar\theta)\; \dot {\bar F}^{(b)}(\tau, \bar\theta)\Big],\nonumber\\
Q_b  & = & \frac {\partial}{\partial\theta}\;\Big[i\;F^{(ab)}(\tau,\theta)\dot F^{(ab)}(\tau,\theta)\Big]\equiv \int \;d\theta\;\Big[i\;F^{(ab)}(\tau,\theta)\dot F^{(ab)}(\tau,\theta)\Big], 
\end{eqnarray}
where we have expressed the BRST charge ($Q_b$) as the total derivative w.r.t. the translational generators
($\partial_\theta$) of the {\it chiral} supermanifold and anti-BRST 
charge ($ Q_{ab}$) has been able to be written as the total derivative w.r.t. to the translational generator
$(\partial_{\bar\theta}$) of the {\it anti-chiral} supermanifold. It is now elementary to note that we have
the following {\it trivial} equalities:
\begin{eqnarray}
&&\partial_\theta\, Q_b = 0, \qquad\qquad \partial_{\bar\theta}\, Q_{ab} = 0 \qquad  \Longleftrightarrow \qquad \partial_\theta^2  
= \partial_{\bar \theta}^2 = 0.
\end{eqnarray}
Thus, we observe that, in contrast to our results in (50), we see that the action of the 
translational generator $\partial_{\theta}$ of the {\it chiral} supermanifold on the BRST charge $(Q_b)$
and action of the derivative ${\partial_{\bar\theta}}$ (i.e. the translational generator along $\bar\theta $-direction
of the {\it anti-chiral} supermanifold) on the anti-BRST charge turn out to produce {\it zero} result, too
(primarily due to the nilpotency property: $\partial_\theta^2 = \partial_{\bar\theta}^2 = 0$).

The consequences of the above observations, within the framework of (anti-)chiral supervariable approach to BRST
formalism, become very transparent and clear when we express the preceding three {\it equations} in the ordinary 
space with the backing of our knowledge of the mappings: $s_b\leftrightarrow  \partial_{\bar\theta},
s_{ab} \leftrightarrow \partial_{\theta}$. It is elementary to check that the expressions 
for the conserved and nilpotent (anti-)BRST charges $(Q_{(a)b}$) can be {\it also} expressed in terms of the (anti-)BRST transformations 
$(s_{(a)b})$ (in a {\it different} form  than (51)) as: 
\begin{eqnarray}
 Q_b = s_{ab}(i\,c\,\dot c), \quad\quad Q_{ab} = s_b (-i\, \bar c \;\dot{\bar c}).
\end{eqnarray}   
In other words, it is quite obvious, from the above equations, that we have the following interesting relationships:
\begin{eqnarray}
&& s_{ab} Q_b = -i\, {\{Q_b, Q_{ab}}\} = 0\quad\Longleftrightarrow \quad s_{ab}^2 = 0,\nonumber\\
&&s_b Q_{ab}  = - i\, {\{Q_{ab}, Q_b}\} = 0\quad\Longleftrightarrow \quad s_b^2 = 0.
\end{eqnarray} 
Thus, we note that the absolute anticommutativity property of the nilpotent (anti-)BRST charges is 
hidden in the nilpotency of the (anti-)BRST symmetry transformations in the {\it ordinary space}
because BRST charge can be expressed as an {\it anti-BRST-exact} quantity and  anti-BRST charge can be 
expressed as the {\it BRST-exact}. We would like to lay emphasis on the fact that 
we have been able to derive equations (51) and (54) because of our knowledge of the (anti-)chiral supervariable
approach to BRST formalism.

We would like to stress on the novel observation that we have made in our 
present investigation. It should be clearly noted that we have  considered {\it only}
the (anti-)chiral super expansions of the supervariables. As a consequence, 
it is {\it not} obvious, at the outset, that the property of the absolute {\it anticommutativity}
of the (anti-)BRST symmetries as well as their corresponding nilpotent and conserved charges 
would be obvious. This is due to fact that our earlier works on ${\cal N} = 2$  supersymmetric 
quantum mechanical models [29-32] show that the symmetry transformations and corresponding  conserved
charges are {\it not}  absolutely anticommuting  even though we have applied the 
(anti-)chiral supervariable approach for the derivation of nilpotent supersymmetry transformations for a  set of  
very interesting ${\cal N} = 2$ supersymmetric quantum mechanical models.
 Thus, our observation of the absolute anticommutativity property of the conserved and nilpotent 
(anti-)BRST charges is a {\it novel} observation. We would like to add that we have {\it now} finally 
established that, in the context of BRST formalism (applied to the gauge and reparameterization invariant theories), 
the absolute anticommutativity of the (anti-)BRST charges ensue {\it despite } the fact
that we consider {\it only} the (anti-)chiral super expansions of the supervariables/superfields
within the framework of (anti-)chiral supervariable/superfield  approach to BRST formalism (where 
the {\it quantum} gauge  (i.e. (anti-)BRST) invariant restrictions on the supervariables/superfields play a decisive role).   
The above observation is crucial and universal.

We capture now the nilpotency of the (anti-)BRST charges $Q_{(a)b}$ that have been quoted in Eq. (17)
for a massless spinning relativistic particle. By the trial and error method, we observe that these charges 
can be expressed in terms of the supervariables that have been obtained after the application of (anti-)BRST invariant
restrictions (cf. Eqs. (37),(47)). In other words, we have the following expression for BRST charge $Q_b$ (cf. Eq. (17)) in terms
of the geometrical quantities on the anti-chiral supermanifold, namely;
\begin{eqnarray}
Q_b  & = & \frac{\partial}{\partial\bar\theta}\,\Big[ \frac {1}{2} \,P_\mu^{(b)}(\tau, \bar\theta) \, X^{\mu(b)}(\tau, \bar\theta) 
+ B^{(b)} (\tau, \bar\theta) \, E^{(b)}(\tau, \bar\theta) \nonumber\\
& + &\tilde \beta ^{(b)}(\tau, \bar\theta) \;\tilde {\bar \beta} ^{(b)}(\tau, \bar\theta)
\, E^{(b)}(\tau, \bar\theta)\Big]\nonumber\\
& \equiv &\int d\,\bar\theta \; \Big[ \frac {1}{2} \,P_\mu^{(b)}(\tau, \bar\theta) \, X^{\mu(b)}(\tau, \bar\theta) 
+ B^{(b)} (\tau, \bar\theta) \, E^{(b)}(\tau, \bar\theta)\nonumber\\
&  + & \tilde \beta ^{(b)}(\tau, \bar\theta) \;\tilde {\bar \beta} ^{(b)}(\tau, \bar\theta)\,
E^{(b)}(\tau, \bar\theta)\Big],
\end{eqnarray}
where {\it all} the supervariables (with  superscript $(b)$) have been derived earlier in Sec. 4. It is clear that 
$\partial_{\bar\theta}\,Q_b = 0$ due to the nilpotency ($\partial_{\bar\theta}^2 = 0$) of the translational generator
$\partial_{\bar\theta}$ along  $\bar\theta$-direction of the (1, 1)-dimensional {\it anti-chiral}
supermanifold on which {\it all} the variables of our theory have been generalized. These  observations become more transparent
and lucid when we 
express (56) in the {\it ordinary} 1D space of the toy model of a massless {\it spinning} relativistic particle because the BRST charge 
can be expressed in the {\it BRST-exact} form as follows
\begin{eqnarray}
&&Q_b  = s_b \Big [ \frac {1}{2}\; p_\mu\,x^\mu + b\,e\, + \beta\,\bar\beta\,e \Big]
\quad\Longleftrightarrow\quad s_b\,Q_b = 0\quad\Longleftrightarrow \quad s_b^2 = 0,\nonumber\\ 
\end{eqnarray} 
which implies that the BRST charge is nilpotent of order two, namely;
\begin{eqnarray}
s_b Q_b  = - i\, {\{Q_b, Q_b}\} = 0\qquad\Longleftrightarrow\qquad Q_b^2 = 0\qquad\Longleftrightarrow\qquad s_b^2 = 0.
\end{eqnarray}
In other words, we observe that the nilpotency ($Q_b^2 = 0$) of the BRST charge $Q_b$ is intimately 
connected with the nilpotency ($s_b^2 = 0$) of the BRST symmetry transformations $s_b$ (cf. Eq. (10)).
Within  the framework  of (anti-)chiral supervariable approach to BRST formalism, we note that the nilpotency 
of BRST charge as well as the  nilpotency of the BRST symmetry transformations is deeply related with 
the nilpotency ($\partial_{\bar\theta}^2 = 0)$ of the translational generator $\partial_{\bar\theta}$
along $\bar\theta$-direction of the {\it anti-chiral} supermanifold.

We dwell a bit now on the off-shell nilpotency of the anti-BRST charge $Q_{ab}$
(cf. Eq. (17)) within the framework of the (anti-)chiral supervariable approach
to BRST formalism. By the method of trial and error and keen observations, 
we note that the anti-BRST charge $(Q_{ab})$ can be written on the 
(1, 1)-dimensional {\it chiral} supermanifold as
\begin{eqnarray}
Q_{ab}  & = & \frac{\partial}{\partial\theta}\,\Big[ \frac {1}{2} \,P_\mu^{(ab)}(\tau, \theta) \, X^{\mu(ab)}(\tau, \theta) 
- \tilde {\bar B}^{(ab)} (\tau, \theta) \, E^{(ab)}(\tau, \theta)\nonumber\\
 & - & \tilde \beta ^{(ab)}(\tau, \theta) \;\tilde {\bar \beta} ^{(ab)}(\tau, \theta)\, E^{(ab)}(\tau, \theta)\Big]\nonumber\\
& \equiv   &\int d\,\theta \; \Big[ \frac {1}{2} \,P_\mu^{(ab)}(\tau, \theta) \, X^{\mu(ab)}(\tau, \theta) 
-\tilde {\bar B}^{(ab)} (\tau, \theta) \, E^{(ab)}(\tau, \theta)\nonumber\\
&  - & \tilde \beta ^{(ab)}(\tau, \theta) \;\tilde {\bar \beta} ^{(ab)}(\tau, \theta)\,
E^{(ab)}(\tau, \theta)\Big],
\end{eqnarray}  
where the supervariables, with superscript $(ab),$ have been explained and derived earlier in Sec. 4.
A close look at the above equation (59) immediately implies that
\begin{eqnarray} 
\partial_\theta \;Q_{ab} = 0\qquad\qquad\Longleftrightarrow \qquad\qquad\partial{_\theta}^2 = 0.  
\end{eqnarray}
The above equations (59) and (60) can be translated into the ordinary space due to our knowledge ($s_{ab}\longleftrightarrow \partial_\theta$) of
the connection between the anti-BRST symmetry transformation $s_{ab}$ and the translational generator $\partial_\theta$
along $\theta$-direction of {\it chiral} (1, 1)-dimensional supermanifold. In the {\it ordinary} 1D space, we have the following:
\begin{eqnarray}
&&Q_{ab}  = s_{ab} \Big [ \frac {1}{2}\; p_\mu\,x^\mu - \bar b\,e\, - \beta\,\bar\beta\,e \Big],
\quad\quad s_{ab}\,Q_{ab} = 0\qquad\Longleftrightarrow \qquad s_{ab}^2 = 0,\nonumber\\ 
&&s_{ab} Q_{ab}  = - i\, {\{Q_{ab}, Q_{ab}}\} = 0\qquad\quad\Longleftrightarrow\quad Q_{ab}^2 = 0\quad\Longleftrightarrow\quad s_{ab}^2 = 0.
\end{eqnarray}
Thus, we conclude that the nilpotency of anti-BRST charge is deeply connected with the nilpotency
($s_{ab}^2 = 0$) of the anti-BRST symmetry transformations $s_{ab}$ which, in turn, is deeply related with 
the nilpotency ($\partial_{\theta}^2 = 0$) of the translational generator $\partial_\theta$ along $\theta$-direction 
of the (1, 1)-dimensional {\it chiral} supermanifold. In other words, the nilpotency properties of $Q_{ab}, s_{ab}$ 
and $\partial_\theta$ are {\it intertwined} together in a beautiful and meaningful manner.

Now we capture the absolute anticommutativity of the conserved  BRST and anti-BRST charges $Q_{(a)b}$ (cf. Eq. (17))
which are off-shell nilpotent of order two (i.e. $Q_{(a)b}^2 = 0$). Towards this objective in mind,
first of all, using the following equations of motion (that have been derived from $L_B$), namely;
\begin{eqnarray}
&&\frac {1}{2}\,p^2  = \dot {\bar b} + 2\,\beta\,\dot {\bar\beta} - 2\,\gamma\,\chi \equiv - \dot b - 2\,\gamma\,\chi - 2\,\bar\beta\dot\beta,\nonumber\\
&&p\cdot\psi = 2\,i\,e\,\gamma + 2\,\bar\beta\dot c - 2\,\beta\dot{\bar c}, 
\end{eqnarray} 
we recast the (anti-)BRST charges as:
\begin{eqnarray}
&&Q_{ab} = (b\,\dot{\bar c} - \dot b\,\bar c) + 2\,i\,e\,\bar\beta\,\gamma + \bar\beta^2\,\dot c - 2\,\bar c\,\gamma\chi
- 2\,\bar c\,\bar\beta\dot\beta + 2\,\beta\,\bar\beta^2\,\chi + 2\, b\,\bar\beta\,\chi,\nonumber\\
&&Q_b = (\dot{\bar b}\,c - \bar b\,\dot c) + 2\,i\;e\,\beta\,\gamma-\beta^2\,\dot{\bar c} - 2\,c\,\gamma\,\chi
+ 2\,c\,\beta\,\dot {\bar\beta} - 2\,\beta^2\,\bar\beta\,\chi
- 2\,\bar b\,\beta\,\chi,
\end{eqnarray}
where we have {\it also} used the CF-type restriction: $b + \bar b + 2\,\beta\,\bar\beta = 0$.
The above charges (which have been derived from the off-shell nilpotent charges $Q_{(a)b}$
(cf. Eq. (17))) can be written in the BRST-{\it exact} and anti-BRST-{\it exact} forms, respectively.
We would like to lay emphasis on the fact that 
 we have {\it stated} these results due to our knowledge of the (anti-)chiral supervariable 
approach to BRST formalism. We elaborate a bit on {\it this} aspect of our statement in the next 
paragraph in a clear and cogent manner.

By the method of trial and error and a keen observations of the supervariable expansions in (47), it can be checked that the BRST charge $Q_b$ can be written, in terms of the supervariables, as 
\begin{eqnarray}
Q_b & = &\frac{\partial}{\partial\theta}\Big[i\, F^{(ab)}(\tau, \theta)\,\dot F^{(ab)}(\tau, \theta) - \tilde\beta^{(ab)}(\tau, \theta)\,
\tilde\beta^{(ab)}(\tau, \theta)\,
E^{(ab)}(\tau, \theta)\nonumber\\
& +  & 2\,i\,F^{(ab)}(\tau, \theta)\,\tilde\beta^{(ab)}(\tau, \theta)\,X^{(ab)}(\tau, \theta)\Big]\nonumber\\
& \equiv & \int\,d\,\theta\; \Big[i\, F^{(ab)}(\tau, \theta)\,\dot F^{(ab)}(\tau, \theta) - \tilde\beta^{(ab)}(\tau, \theta)\,
\tilde\beta^{(ab)}(\tau, \theta)\,
E^{(ab)}(\tau, \theta)\nonumber\\
& +  & 2\,i\,F^{(ab)}(\tau, \theta)\,\tilde\beta^{(ab)}(\tau, \theta)\,X^{(ab)}(\tau, \theta)\Big],
\end{eqnarray} 
where the supervariables with superscript $(ab)$ have been written in Eq. (47). It can be explicitly 
checked that the above expression (in the supervariable approach) yields the BRST charge given in Eq. (63).
The absolute anticommutativity of $Q_b$ {\it with} $Q_{ab}$ becomes transparent when we express the above 
charge in the ordinary 1D space in terms of the anti-BRST symmetry transformation $s_{ab}$ 
(with $\partial_\theta\longleftrightarrow s_{ab}$), namely;
\begin{eqnarray}
Q_b = s_{ab}\Big[ i\, c\,\dot c - \beta^2\,e + 2\, i\, c\, \beta\,\chi\Big].
\end{eqnarray}  
The above equation implies the following in a straightforward manner:
\begin{eqnarray}
s_{ab}\,Q_b = -\,i\,{\{Q_b, Q_{ab}}\} = 0\qquad\qquad\Longleftrightarrow \qquad\qquad s_{ab}^2 = 0.
\end{eqnarray}
Thus, we note that the absolute anticommutativity of the BRST charge {\it with} anti-BRST charge is 
encoded in the nilpotency of anti-BRST symmetry transformations (i.e. $s_{ab}^2 = 0$). This becomes 
evident from the anti-BRST-{\it exact} form quoted in Eq. (65). Furthermore a close look at Eq. (64)
makes it clear that $\partial_\theta Q_b  = 0$ due to the nilpotency ($\partial_\theta ^{2} = 0$) of the 
translational generator $\partial_\theta$ along $\theta$-direction of the {\it chiral} supermanifold.
Hence, we draw the conclusion that the {\it absolute anticommutativity} property (cf. Eq. (65))
is deeply connected with the nilpotency property ($s_{ab}^2 = 0, \partial_{\theta}^2 = 0$) associated with 
anti-BRST symmety transformations $s_{ab}$ and translational generator $\partial_\theta$.

We are now in the position to concentrate on the expression for the anti-BRST charge $Q_{ab}$ that has been quoted in (63).
We can capture this expression in the terminology of (anti-)chiral supervariable approach 
to BRST formalism on (1, 1)-dimensional (anti-)chiral supermanifolds. By the trial and error method 
and keen observation  of the expansions in (37), it can be seen that the expression for $Q_{ab}$ can be written as:
\begin{eqnarray}
Q_{ab} & = &\frac{\partial}{\partial\bar\theta}\,\Big[ \tilde\beta^{(b)}(\tau, \bar\theta)\,
\tilde\beta^{(b)}(\tau, \bar\theta)\,
E^{(b)}(\tau, \bar\theta)- i\, \bar F^{(b)}(\tau, \bar\theta)\,\dot {\bar F}^{(b)}(\tau, \bar\theta) \nonumber\\
& -  & 2\,i\,\bar F^{(b)}(\tau, \bar\theta)\,\tilde{\bar\beta}^{(b)}(\tau, \bar\theta)\,X^{(b)}(\tau, \bar\theta)\Big]\nonumber\\
& \equiv & \int\,d\,\bar\theta\; \Big[ \tilde\beta^{(b)}(\tau, \bar\theta)\,
\tilde\beta^{(b)}(\tau, \bar\theta)\,
E^{(b)}(\tau, \bar\theta)- i\, \bar F^{(b)}(\tau, \bar\theta)\,\dot {\bar F}^{(b)}(\tau, \bar\theta) \nonumber\\
& -  & 2\,i\,\bar F^{(b)}(\tau, \bar\theta)\,\tilde{\bar\beta}^{(b)}(\tau, \bar\theta)\,X^{(b)}(\tau, \bar\theta)\Big],
\end{eqnarray} 
where the supervariables with superscript $(b)$ have been derived after application of BRST 
invariant restrictions in (37). It is straightforward to note, from the above equation (67), 
that we have the following relationships:
\begin{eqnarray}
&&\partial_{\bar\theta}\, Q_{ab} = 0\qquad\qquad\Longleftrightarrow\qquad\qquad  \partial_{\bar\theta}^2 = 0.
\end{eqnarray}
The absolute anticommutativity of the anti-BRST charge ($Q_{ab}$) {\it with}  the BRST charge $(Q_b)$
can be expressed in a very transparent manner due to our understanding of $s_b\longleftrightarrow \partial_{\bar\theta}$.
Thus, in the ordinary 1D space, we have the following: 
\begin{eqnarray}
Q_{ab} = s_b\Big[  \bar\beta^2\,e - i\, \bar c\,\dot {\bar c}   - 2\, i\,\bar c\, \bar\beta\,\chi\Big].
\end{eqnarray}
From the above equation, it is straightforward to note that we have 
\begin{eqnarray}
s_b\,Q_{ab} =  - i\,{\{Q_{ab}, Q_b}\} = s_b^2\,\Big[  \bar\beta^2\,e - i\, \bar c\,\dot {\bar c}   - 2\, i\,\bar c\, \bar\beta\,\chi\Big] = 0,
\end{eqnarray} 
which demonstrates that the absolute anticommutativity of the anti-BRST charge {\it with} BRST charge is ultimately
connected with the nilpotency ($s_b^2 = 0$) of the BRST transformations $s_b$ which, in turn, is connected with the nilpotency
($\partial_{\bar\theta}^2 = 0$) of the translational generator $\partial_{\bar\theta}$ along 
the Grassmannian  $\bar\theta$-direction of the {\it anti-chiral}
supermanifold (parameterized by $\tau$ and $\bar\theta$).

\section{Summary and Conclusions}

n our present endeavor, we have discussed two toy models (i.e. a {\it free} massive $scalar$ relativistic particle and
a {\it free} $spinning$ relativistic particle which are the examples of a set of reparameterization invariant theories)  within the framework 
of (anti-)chiral supervariable approach to BRST formalism and established that the {\it absolute anticommutativity} of
the conserved and off-shell nilpotent  (anti-)BRST charges is satisfied  {\it despite} the fact that we have taken into consideration
{\it only} the (anti-)chiral super expansions of the supervariables on a set of suitably chosen (1, 1)-dimensional 
(anti-)chiral supermanifolds. Our results, once again, have  demonstrated that the simple (but intuitive and beautiful)
 (anti-)chiral supervariable/superfield approach (i) to  the {\it gauge } as well as the {\it reparameterization} invariant theories
 is good enough to capture {\it two} of the central mathematical properties (i.e. nilpotency  and
absolute anticommutativity) associated with the (anti-)BRST symmerties and corresponding
 conserved and  charges, and (ii) to provide the geometrical meanings (see, e.g. [17-22]) to the symmetry transformations 
and corresponding conserved charges.

 In the (anti-)chiral supervariable approach to BRST formalism, the key  role is played by the {\it quantum}
gauge [i.e. (anti-)BRST] invariant restrictions  on the (anti-)chiral supervariables that lead to the derivation of 
(anti-)BRST symmetries. Furthermore, we observe that the conserved and nilpotent charges of the theory could be expressed in terms of    
the supervariables  (that are obtained after the application  of the (anti-)BRST invariant restrictions) and some other geometrical  quantities
(e.g. Grassmannian derivative and differentials) that are defined on the suitably chosen (1, 1)-dimensional (anti-)chiral 
 supermanifolds on which the  reparameterization invariant theories  are generalized. Ultimately, this exercise leads to the BRST and anti-BRST
{\it exact} forms of the expressions for the conserved (anti-)BRST charges and this, in turn, produces the nilpotency and 
absolute anticommutativity  of the  BRST and anti-BRST charges within the framework of (anti-)chiral supervariable approach to BRST
formalism. The observation of the absolute anticommutativity property of the (anti-)BRST charges is a completely {\it novel} result
in our {\it present} endeavor (and earlier works [17-22]) in view of the fact that {\it only} the (anti-)chiral super expansions 
of the supervariables have been taken into account.

In our present endeavor, we have applied our (anti-)chiral supervariable approach to 
{\it simple} toy models of an ordinary $scalar$ relativistic particle and a $spinning$ relativistic particle 
and have shown the validity of  absolute anticommutativity of the (anti-)BRST charges where the CF-type 
restriction plays a key role (in the context of supersymmetric  spinning relativistic
particle). The crucial point to be noted is the (non-)existence of  CF-type restriction
in {\it both} the toy models that have been taken into account in our present investigation.
As a consequence of the above observations, we see that the Lagrangian (1) for a scalar 
relativistic particle respects {\it both} BRST and anti-BRST symmetries. However,
there is an existence of a coupled (but equivalent) Lagrangians (cf. Eq. (6)) for the 
description of a free spinning relativistic particle. We note that $L_B$ and  $L_{\bar B}$ (cf. Eq. (6))
 have {\it perfect} BRST and anti-BRST symmetries (cf. Eqs. (11),(12)), respectively,
but when one applies the anti-BRST transformations on $L_B$ {\it and} BRST symmetry transformations on    $L_{\bar B}$,
these Lagrangians transform to the total derivatives {\it plus} terms that vanish due to the CF-restriction [23].

As far as the off-shell nilpotency and absolute anticommutativity of the conserved  BRST and anti-BRST charges are concerned, we observe 
that the off-shell nilpotency of BRST charge is connected with the nilpotency ($\partial_{\bar\theta}^2 = 0$) of the translational generator ($
\partial_{\bar\theta}$) along $\bar\theta$-direction of (1, 1)-dimensional {\it anti-chiral} supermanifold {\it as well as} 
the nilpotency ($s_b^{2} = 0$) of the BRST symmetry transformations ($s_b$). In exactly similar fashion, the off-shell nilpotency 
of anti-BRST charge is connected with the nilpotency properties  $(\partial_{\theta}^2= 0, s_{ab}^2 = 0)$ associated with the translational
generator ($\partial_\theta$)
 along $\theta$-direction of the {\it chiral} supermanifold {\it and} anti-BRST symmetry transformations ($s_{ab}$). On the contrary, we observe
that the { \it absolute anticommutativity} of the BRST charge {\it with} anti-BRST charge is connected with 
the nilpotency properties ($\partial_{\theta}^2 = 0, s_{ab}^2 = 0)$ of $\partial_\theta$ and $s_{ab}$ (cf. Eqs. (64),(65)).
Further, we note that the {\it absolute anticommutativity} of the anti-BRST charge {\it with} BRST charge 
is connected with the nilpotency properties ($\partial_{\bar\theta}^2 = 0, s_b ^2  = 0$) of $\partial_{\bar\theta}$ and $s_b$ (cf. Eqs. (68),(69)).
These latter observations, connected with the property of absolute anticommutativity, are completely  {\it novel} results.
In fact, this is one of the highlights of our present investigation where we have proven the absolute anticommutativity property of the conserved charges
{\it despite } the fact that we have taken into consideration {\it only} the (anti-)chiral super expansions for the supervariables.

It would be a challenging problem for us to apply our method of derivation  of the nilpotent (anti-)BRST 
symmetries to some {\it physical} problems of interest in the  physical {\it four} (3+1)-dimensions 
of spacetime. It would be also interesting to prove the absolute anticommutativity,
nilpotency, etc., of the 4D field theoretic and/or  supersymmetric models of physical interest
that respect the reparameterization invariance. Such theories are, of course, the gravitational as well as (super)string/supergravity
theories where the reparameterization invariance  plays a decisive role. 
We plan to pursue {\it all} the above cited problems  in the forthcoming years through our future publications [33].\\

\noindent
{\bf Acknowledgements}:
Financial supports from the BHU-fellowship and DST-INSPIRE fellowship are gratefully
acknowledged by S. Kumar and B. Chauhan, respectively. The present work has been carried out
under the above fellowships.\\

\begin{center}
{\bf Appendix A: On the Nilpotent (Anti-)BRST Charges  }\\
\end{center}

{\vskip 0.3cm}

\vspace{0.5 cm}

\noindent
We comment here on the derivation of {\it equivalent} expressions for the {\it conserved}
charges (cf. Eq. (17)) and their nilpotency properties by using the idea of continuous symmetry transformations
and their generators. First of all, we note that the expressions for the (anti-)BRST 
charges (that are derived {\it directly} by using the Noether  theorem) are 
\[
Q_{ab}^{(1)}   = \frac {1}{2}\;\bar c\, p^2 -\bar b \,\dot{\bar c} + \bar\beta \,(p\cdot\psi) 
- \bar \beta^2\, \dot c  - 2 \beta \,\bar\beta^2\,  \,\chi  - 2 \,\bar b\, \bar \beta\,\chi,\]
\[Q_b^{(1)}   =  \frac {1}{2}\, c\, p^2 + b \,\dot c + \beta \,(p\cdot\psi) 
+ \beta^2\, \dot{\bar c} + 2 \,\beta^2\,\bar\beta   \,\chi + 2 \,b\, \beta\,\chi,\eqno (A.1)\]
where the continuous symmetry transformations (9) and (10) have been used in the following explicit expressions for the
 Noether conserved (anti-)BRST charges in the context of a 1D supersymmetric toy model of a $spinning$ 
relativistic particle (cf. Eqs. (11),(12)):
\[ 
Q_{ab}^{(1)} = s_{ab} \phi_i \;\Big(\frac{\partial L_{\bar B}}{\partial \phi_i}\Big) - \frac {1}{2}\; \bar c\,p^2 
- \frac {1}{2}\;\bar\beta \;(p\cdot\psi) +\bar b\;(\dot {\bar c} + 2\,\bar\beta\,\chi),\]
\[Q_b^{(1)} = s_b \phi_i \;\Big(\frac{\partial L_B}{\partial \phi_i}\Big) - \frac {1}{2}\; c\,p^2 - \frac {1}{2}\;\beta\; (p\cdot\psi) - b\;(\dot c + 2\,\beta\,\chi).\eqno (A.2)\]
In the above, we have $\phi_i(\equiv x_\mu, e, \bar c, c, b,  \bar b, \psi_\mu, p_\mu, \beta, \bar\beta, \gamma)$ as the 
generic variable of the theory. If we apply directly $s_{(a)b}$ on the above Noether charges (cf. (A.1)), we obtain:
\[
s_{ab}\,Q_{ab}^{(1)} = \frac {1}{2}\;i\,\bar\beta^2\,p^2 - i\,\bar \beta^2\,\dot {\bar b}+ 2\,i\,\bar \beta^2\,\gamma\,\chi - 2\,i\,
\bar\beta^2\,\beta\,\dot{\bar\beta},\]
\[s_b\,Q_b^{(1)} = \frac {1}{2}\;i\,\beta^2\,p^2 + i\, \beta^2\,\dot b + 2\,i\, \beta^2\,\gamma\,\chi + 2\,i\,\beta^2\,\bar\beta\,\dot\beta.\eqno (A.3)\]
The above equations (with the inputs $s_b Q_b^{(1)}  = -i\, {\{Q_b^{(1)}, Q_b^{(1)}}\}, 
\; s_{ab}Q_{ab}^{(1)} = - i {\{Q_{ab}^{(1)}, Q_{ab}^{(1)}}\}$)
demonstrate that the conserved charges $Q_{(a)b}^{(1)}$
(cf. (A.1)) are {\it not } nilpotent of order two {\it unless} we apply
 the following equations of motion (derived from $L_B $ and $L_{\bar B}$):
 \[\dot b = -\,\frac {1}{2}\;p^2 - 2\;\gamma\,\chi - 2\,\bar\beta\,\dot\beta,\qquad\quad \dot {\bar b} = \,\frac {1}{2}\;p^2 + 2\,\gamma\,\chi - 2\,\beta\,\dot {\bar \beta}.\eqno (A.4)\]
Thus, we note that the (anti-)BRST charges, derived {\it directly} from the Noether theorem, are {\it not} off-shell
nilpotent {\it despite} the fact that the (anti-)BRST symmetries (quoted in Eqs. (9), (10)) are themselves off-shell nilpotent.
This is the reason   that the (anti-)BRST charges (A.1) have been recast in a different form using the EOMs.
These appropriate off-shell nilpotent charges have been quoted in Eq. (17) and have been denoted by $Q_b^{(2)} \equiv Q_b$ 
and $Q_{ab}^{(2)} \equiv Q_{ab}$.

We end this Appendix with the remark that there are other {\it equivalent} expressions for the conserved charges 
$Q_{(a)b}$ that are {\it on-shell} nilpotent. These are listed below:
\[Q_{ab}^{(3)} \equiv  \dot {\bar b}\;\bar c - \bar b \;\dot {\bar c} - 2\,\bar c\gamma\,\chi + 2\,\bar c\,\beta\,\dot{\bar \beta} + \bar\beta (p\cdot\psi) - \bar \beta^2\, \dot c  - 2 \beta \,\bar\beta^2\,  \,\chi  - 2\bar b\, \bar \beta\,\chi,\]
\[Q_{ab}^{(4)}\equiv  \dot {\bar b}\bar c - \bar b\,\dot {\bar c}  +\frac {1}{2}\bar \beta\,(p\cdot\psi) - 2\,\bar c\,\gamma\,\chi\, - 2 \bar c\,\beta\,\dot{\bar\beta}
- \bar\beta\,\beta\, \dot {\bar c} 
- i\, \bar\beta\,e\,\gamma
  -2 \bar\beta^2\,\beta   \,\chi - 2 \bar b\, \bar\beta\,\chi,\]
\[Q_b^{(3)}\equiv  b\dot c - \dot b c - 2\,c\gamma\,\chi - 2\,c\bar\beta\dot \beta + \beta\; (p\cdot\psi) + \beta^2\, \dot{\bar c} +2 \beta^2\,\bar\beta   \,\chi + 2 b\, \beta\,\chi,\]
\[Q_{b}^{(4)}\equiv  b\dot c - \dot b c +\frac {1}{2} \beta\,(p\cdot\psi) - 2\,c\,\gamma\,\chi\, - 2 c\,\bar\beta\,\dot\beta
 + \beta\,\bar\beta\, \dot c  + i\, \beta\,e\,\gamma 
+2 \beta^2\,\bar\beta   \,\chi + 2 b\, \beta\,\chi.\eqno (A. 5)\]
We can check that the above charges are nilpotent (i.e. $s_b Q_b^{(3, 4)}  = 0$ and $s_{ab} Q_{ab}^{(3,4)} = 0$)
{\it only} when we use the EOM (A.4). We note that the above charges 
 $ Q_b^{(3, 4)} $ and $Q_{ab}^{(3,4)}$ have been derived from (A.1) by using the appropriate EOMs that are listed in Eqs. (14) and (15).
Thus, we draw the conclusion that the conserved (anti-)BRST charges $Q_{(a)b}^{(1, 3, 4)}$ are nilpotent {\it only} when we use the 
EOMs. However, the conserved (anti-)BRST charges $Q_{(a)b}^{(2)}$ are nilpotent {\it without} any use of EOMs.
This is why, we have chosen {\it these} charges (in Eq. (17)) 
whose {\it absolute anticommutativity} property has been derived and discussed in Sec. 5. \\

\begin{center}
{\bf Appendix B: On the Absolute Anticommutativity Property}\\
\end{center}

{\vskip 0.3cm}

\noindent
In this Appendix, we dwell a bit on the property of absolute anticommuatativity of the
(anti-)BRST charges that have been discussed in Sec. 2 (and Appendix A) in the context of
1D toy model of a  massless $spinning$ relativistic particle. When we apply the anti-BRST symmetry transformations
on BRST charge $(Q_b)$ and BRST symmetry transformation on anti-BRST charge $(Q_{ab})$ (cf. Eq. (17)),
we obtain the following: 
\[s_{ab}\;Q_b = \frac{1}{2}\;i\;p^2\,(\bar b + \beta\,\bar\beta) + i\; (b + \beta\,\bar\beta)\dot{\bar b} - 2i\;(b + \beta\,\bar\beta)\gamma\chi\]
\[ + 2i\;\beta\;(b + \beta\,\bar\beta)\dot{\bar\beta}
  -\;i\;\bar\beta\,\gamma\dot c - i\;\bar\beta\,\gamma\,\dot{\bar c} - \frac{i}{2}\;\gamma\,(p\cdot\psi),\]
\[s_b\;Q_{ab} = \frac{1}{2}\;i\;p^2\,( b + \beta\,\bar\beta) - i\; (\bar b + \beta\,\bar\beta)\dot b - 2i\;(\bar b + \beta\,\bar\beta)\gamma\chi\]
\[ 2i\;\bar\beta\;(\bar b + \beta\,\bar\beta)\dot\beta
   + i\;\beta\,\gamma\,\dot{\bar c} + i\;\bar\beta\,\dot c\,\gamma + \frac{i}{2}\;\gamma\,(p\cdot\psi).\eqno (B.1)\] 
Using the CF-type restriction ($b + \bar b + 2\,\beta\bar\beta = 0 $) which implies that $\beta\bar\beta + b = -\,(\bar b  + \beta\,\bar\beta$),
we obtain the following:
\[s_{ab}\;Q_b = i\;( b + \beta\bar\beta)\;\Big[\dot{\bar b}  - \frac{p^2}{2} + 2\beta\dot{\bar\beta}- 2\gamma\chi\Big] - i\,\gamma\;
\Big[\frac{1}{2}\,(p\cdot\psi) - \bar\beta\,\dot c + \beta\,\dot{\bar c}\Big],\]
\[s_b\;Q_{ab} = -\,i\;( \bar b + \beta\bar\beta)\;\Big[\dot b + \frac{p^2}{2} + 2\beta\dot{\bar\beta}+ 2\gamma\chi\Big]
+ i\,\gamma\;
\Big[\frac{1}{2}\,(p\cdot\psi) - \bar\beta\,\dot c + \beta\,\dot{\bar c}\Big]. \eqno (B.2)\]
It is crystal clear that if we use the EOMs (A.4) and $\frac{1}{2} (p\cdot\psi) + \beta\,\dot{\bar c} - \bar\beta\dot c = i\,e\,\gamma,$
we obtain the absolute anticommutativity (${\{Q_b, Q_{ab}}\} = 0$) of the charges
 due to the basic relationship between the continuous symmetry transformations
and the conserved and nilpotent charges as the generators in: $s_b Q_{ab}  = -\,i\,\{Q_{ab}, Q_b \} = 0$ and $s_{ab} Q_b  = -\,i\,\{Q_b, Q_{ab}\} = 0$.

To prove the absolute anticommutativity of the (anti-)BRST charges $(Q_{(a)b}^{(1)}$) that have been derived 
{\it directly} by using the Noether theorem, we note that:
\[s_{ab}\, Q_b ^{(1)} = \frac {1}{2} \,i\,p^2 \Big [ \bar b + 2\, \beta\,\bar\beta\Big] + i\, b \Big [ \dot {\bar b} + 2\,\beta\,\dot{\bar\beta}
- 2\,\gamma\,\chi\Big]
 i\, \gamma\,\Big[ (p\cdot\psi) - 2\, \bar\beta\,\dot c + 2\,\beta\dot {\bar c}\Big],\]
\[s_b\, Q_{ab} ^{(1)} = \frac {1}{2} \,i\,p^2 \Big [  b + 2\, \beta\,\bar\beta\Big] - i\, \bar b \Big [ \dot  b + 2\,\dot\beta\,\bar\beta
+ 2\,\gamma\,\chi\Big]
+  i\, \gamma\,\Big[ (p\cdot\psi) + 2\, \beta\,\dot {\bar c} - 2\,\bar\beta\dot  c\Big].\eqno (B.3)\]
Using the equations of motion (A.4), CF-type restriction ($b + \bar b + 2\,\beta\,\bar\beta = 0)$ and ($p\cdot\psi) - 2\,\bar\beta\,\dot c
+ 2\,\beta\,\dot {\bar c}  = 2\,i\,e\,\gamma$, it can be seen that
\[s_{ab}Q_b^{(1)} = -\, \frac {1}{2}\; i\;b\,p^2 +   \frac {1}{2}\; i\;\,b\,p^2 = 0\quad \Longleftrightarrow \quad  - i\,\,{\{Q_b^{(1)}, Q_{ab}^{(1)}}\} = 0,\]
\[s_b\,Q_{ab}^{(1)} = -\, \frac {1}{2}\; i\;\bar b\,p^2 +   \frac {1}{2}\; i\;\,\bar b\,p^2 = 0\;\Longleftrightarrow \; - i\,\,{\{Q_{ab}^{(1)}, Q_b^{(1)}}\} = 0,\eqno (B.4)\]
 where we have taken into account the key relationship between the symmetry generators and continuous symmetry 
transformations  so that $s_b\,Q_{ab}^{(1)} =
-\,i\,{\{Q_{ab}^{(1)}, Q_b^{(1)}}\} = 0$ and $s_{ab}\,Q_b^{(1)} =
-\,i\,{\{Q_b^{(1)}, Q_{ab}^{(1)}}\} = 0$.
In other words, we have proven the absolute anticommutativity of the (anti-)BRST charges $(Q_{(a)b}^{(1)})$ that are derived 
{\it directly} by using the Noether theorem. However, we note that we have used the EOMs (15) and (14) {\it in addition} to the 
CF-type restriction for this proof. We draw the conclusion that we have to find out the appropriate form of the (anti-)BRST 
charges which could be written {\it precisely} in the BRST-exact and anti-BRST-{\it exact} forms so that the absolute anticommutativity
could be proven {\it only} by using  the CF-type restriction. This has been 
precisely and elegantly achieved (cf. Eqs. (65), (69)) in the main body of our text (cf. Sec. 5).\\

\begin{center}
{\bf Appendix C:  On   the (Anti-)BRST Invariance }\\
\end{center}

{\vskip 0.3cm}

\noindent
In this Appendix, we concisely discuss the (anti-)BRST invariance of the Lagrangians $L_{\bar B}$ and $L_B$
within the framework of (anti-)chiral supervariable approach to BRST formalism. In this context, first of all,
we note that the {\it starting} Lagrangian ($L_0$) for our 1D toy model of spinning relativistic particle (cf. Eq. (7)) remains 
invariant under the (anti-)BRST symmetry transformations (cf. Eqs. (9) and (10)). It can be explicitly
checked that:
\[s_b L_0 = \frac {d}{d\tau} \Big [ \frac {1}{2}\,    c\,p^2 + \frac {1}{2}\, \beta \,(p\cdot\psi) \Big], \]
\[s_{ab} L_0 = \frac {d}{d\tau} \Big [ \frac {1}{2}\,    \bar c\,p^2 + \frac {1}{2}\, \bar\beta\; (p\cdot\psi) \Big].\eqno (C.1)\]
The above invariance can be captured within the framework of (anti-)chiral supervariable approach to BRST formalism. 
Towards this goal in mind, we generalize the Lagrangian $L_0$ to its counterpart {\it super} Lagrangians  as
\[L_0\longrightarrow \tilde L_0^{(ab)}   =  p_\mu(\tau)\,\dot X^{\mu{(ab)}}(\tau,\theta)
 - \frac {1}{2} E^{(ab)}(\tau,\theta)\, p^2(\tau)
+  \frac {i}{2}\, \Psi_{\mu}^{(ab)}(\tau,\theta)\,\dot \Psi^{\mu(ab)}(\tau,\theta)\]
\[ + \; i\, X^{(ab)}(\tau,\theta)\, p_\mu(\tau)\, \Psi^{\mu(ab)} (\tau,\theta),\eqno (C.2)\]
\[L_0\longrightarrow \tilde L_0^{(b)}  =  p_\mu(\tau)\,\dot X^{\mu{(b)}}(\tau,\bar\theta) 
- \frac {1}{2} E^{(b)}(\tau,\bar\theta)\, p^2(\tau)\]
\[+  \frac {i}{2}\, \Psi_{\mu}^{(b)}(\tau,\bar\theta)\,\dot \Psi^{\mu(b)}(\tau,\bar\theta)
+ \; i\, X^{(b)}(\tau,\bar\theta)\, p_\mu(\tau)\, \Psi^{\mu(b)} (\tau,\bar\theta),\eqno (C.3)\]
 where we have taken into account $P_\mu^{(b)}(\tau, \bar\theta) = p_\mu (\tau)$ and $P_\mu^{(ab)}(\tau, \theta) = p_\mu (\tau)$ 
because $ p_\mu(\tau)$ is an (anti-)BRST invariant quantity (i.e. $s_{(a)b}\,p_\mu(\tau) = 0)$.
Now it is an elementary exercise to observe that we have the following:
\[\frac {\partial}{\partial\theta}\tilde L_0^{(ab)} = \frac {d}{d\tau}\Big [ \frac {1}{2}\,    \bar c\,p^2 + \frac {1}{2}\, \bar\beta (p\cdot\psi) \Big],\]
\[\frac {\partial}{\partial\bar\theta}\tilde L_0^{(b)} = \frac {d}{d\tau}\Big [ \frac {1}{2}\,    c\,p^2 + \frac {1}{2}\, \beta (p\cdot\psi) \Big].\eqno (C.4)\]
Taking into account the mapping: $s_b \longleftrightarrow \partial_{\bar\theta}, s_{ab} \longleftrightarrow \partial_\theta$,
we can translate the above equation (C.4) into the ordinary 1D space of the toy model of a {\it spinning} relativistic particle
which, ultimately, boils down to the results that have been quoted in (C.1). We would like to point out that 
the superscripts $(ab)$ and $(b)$ on the $super$ Lagrangians (C.2) and (C.3) denote the fact that these Lagrangians
have been expressed in terms of the supervariables that have been obtained after the anti-BRST and BRST invariant restrictions.

We would like to capture now the  (anti-)BRST invariance of the gauge-fixing 
and Faddeev-Popove (FP) ghost terms (from the Lagrangian $L_{\bar B}$ and $L_B$)
within the framework of (anti-)chiral supervariable approach to BRST formalism. Towards this objective in mind, we have to
focus {\it first} on $L_B$ where we take into account $\tilde\beta ^{(b)} (\tau, \bar\theta) = \beta(\tau), 
G^{(b)}(\tau, \bar\theta)  = \gamma(\tau), \tilde B^{(b)} (\tau, \bar\theta) = b(\tau)$. With these as inputs,
we can generalize the gauge-fixing and FP-ghost terms (of $L_B$) in terms of appropriate supervariables  as
\[\tilde L_B^{(gf)} + \tilde L_B^{(fp)} =   b(\tau) \,\dot E^{(b)}(\tau, \bar\theta) + b(\tau)\,\Big[b(\tau) 
 + 2\,\beta(\tau)\tilde{\bar \beta}^{(b)}(\tau, \bar\theta)\Big]\]
\[~~~~~~~~~~~~~ -  \,\,i\,\dot{\bar F}^{(b)}(\tau, \bar\theta)
\Big[\dot F^{(b)}(\tau, \bar\theta) + 2\,\beta(\tau) \,X^{(b)}(\tau, \bar\theta)\Big]\]
\[~~~~~~~~~~~~~~~~~~~~~~~~~~~~~~~+  2\,i\,\tilde{\bar \beta} ^{(b)}(\tau, \bar\theta)\,
\dot F^{(b)} (\tau, \bar\theta)\,X^{(b)}(\tau, \bar\theta)+ 2\,\beta(\tau)\, \gamma(\tau)\,\bar F^{(b)}(\tau, \bar\theta)\]
\[~~~~~~~~~~~~~~~~~~-\,2\, E^{(b)}(\tau, \bar\theta)\Big[\gamma(\tau)\,
X^{(b)}(\tau, \bar\theta) + \tilde{\bar\beta}^{(b)}(\tau, \bar\theta)\,
\dot\beta(\tau)\Big] \]
\[~~~~~~~~~~~~~~~~~~~+ \beta ^2 (\tau) \,\tilde{\bar \beta} ^{2(b)}(\tau, \bar\theta)
 + 2\,\tilde{\bar \beta} ^{(b)}(\tau, \bar\theta)\,F^{(b)}(\tau, \bar\theta)\;\gamma (\tau),\eqno (C.5)\]
where the superscript $(b)$ denotes the fact that we have taken into account the super expansions (37)
for the supervariables of our theory. 
It can be checked explicitly that we have the following:
\[\frac{\partial}{\partial\bar\theta}\Big[\tilde L_B^{(gf)} + \tilde L_B^{(fp)}\Big] =
 \frac {d}{d \tau}\, \Big[b\,(\dot c + 2\beta\,\chi)\Big].\eqno (C.6)\]
In exactly similar fashion, we can capture the result of the application of off-shell
nilpotent anti-BRST symmetry transformations on the gauge-fixing and Faddeev-Popov ghost terms of the Lagrangian $L_{\bar B}$.
Towards this goal in mind, first of all, we take into account $G^{(ab)}(\tau, \theta) = \gamma (\tau), \tilde {\bar B}^{(ab)}(\tau, \theta)
 = \bar b(\tau),  \tilde{\bar\beta}^{(ab)}(\tau, \theta)
 = \bar\beta(\tau).$
With these as inputs, we have the following  form of the super Lagrangian for the gauge-fixing and FP ghost terms
\[\tilde L_{\bar B}^{(gf)} + \tilde L_{\bar B}^{(fp)} = -\,\bar b(\tau) \,\dot E^{(ab)}(\tau, \theta) + \bar b(\tau)\,\Big[\bar b(\tau) 
 + 2\,\bar\beta(\tau) \, \tilde\beta^{(ab)}(\tau, \theta)\Big]\]
\[-\, i\,\dot{\bar F}^{(ab)}(\tau, \theta)
\Big[\dot F^{(ab)}(\tau, \theta) + 2\,\tilde\beta^{(ab)}(\tau, \theta) \,X^{(ab)}(\tau, \theta)\Big]\]
\[+ 2\,i\,{\bar \beta}(\tau)
\dot F^{(ab)} (\tau, \theta)\,X^{(ab)}(\tau, \theta)+ 2\,\tilde\beta^{(ab)}(\tau, \theta)\, \gamma(\tau)\,\bar F^{(ab)}(\tau, \theta)\]
\[-\,2\, E^{(ab)}(\tau, \theta)\Big[\gamma(\tau)\,
X^{(ab)}(\tau, \theta) - \tilde{\beta}^{(ab)}(\tau, \theta)\,
\dot{\bar\beta}(\tau)\Big] \]
\[+ {\bar\beta}^2 (\tau) \,\tilde\beta ^{2(ab)}(\tau, \theta)
 + 2\,\tilde\beta^{(ab)}(\tau, \theta)\,F^{(ab)}(\tau, \theta)\;\gamma (\tau),\eqno (C.7)\]
where {\it all} the supervariables with superscript $(ab)$ have been derived earlier in terms of the
anti-BRST symmetry transformations $s_{ab}$ (cf. Eq. (47)). Finally, we note that\footnote{A close look
at the equations (C.6) and  (C.8) demonstrates that the gauge-fixing and Faddeev-Popov ghost terms of {\it both} the
Lagrangians $L_B$ and $L_{\bar B}$ are BRST and anti-BRST invariant, respectively.} 
\[\frac{\partial}{\partial\theta}\Big[\tilde L_{\bar B}^{(gf)} + \tilde L_{\bar B}^{(fp)}\Big] =
 \frac {d}{d \tau}\, \Big[-\,\bar b\,(\dot {\bar c} + 2\bar\beta\,\chi)\Big].\eqno (C.8)\]
We end our discussion on the (anti-)chiral supervariable approach to BRST formalism in the 
context of the 1D toy model of a $spinning$ relativistic particle by the observation that {\it sum} 
of our results in (C.8) and (C.4) {\it as well as} (C.6) and (C.4) produce the following
\[\frac{\partial}{\partial\theta}\Big[\tilde L_0 + \tilde L_{\bar B}^{(gf)} + \tilde L_{\bar B}^{(fp)}\Big] =
 \frac {d}{d \tau}\, \Big[ \frac {1}{2}\,    \bar c\,p^2 + \frac {1}{2}\, \bar\beta (p\cdot\psi)-\,\bar b\,(\dot {\bar c} 
+ 2\bar\beta\,\chi)\Big],\]
\[\frac{\partial}{\partial\bar\theta}\Big[\tilde L_0+\tilde L_B^{(gf)} + \tilde L_B^{(fp)}\Big] =
 \frac {d}{d \tau}\, \Big[\frac {1}{2}\,    c\,p^2 + \frac {1}{2}\, \beta (p\cdot\psi)
+ b\,(\dot c + 2\beta\,\chi)\Big],\eqno (C.9)\]
which are nothing but the results that have been quoted in Eqs. (12) and (11) for the 1D toy model of a {\it spinning}
relativistic particle in the {\it ordinary} space. Thus, we have captured the (anti-)BRST invariance of the {\it action integrals}  corresponding 
to the Lagrangians $L_{\bar B}$ and $L_B$ for the 1D toy model of a free spinning (i.e supersymmetric) relativistic particle.

We end this Appendix with a concise discussion on the (anti-)BRST invariance (cf. Eq. (3)) of the Lagrangian $L_b$ (cf. Eq. (1))
for a free {\it scalar} relativistic particle. Taking the help of our super expansions in (22) and (27),
it can be seen that the Lagrangian $L_b$ can be generalized to its counterparts {\it super} Lagrangians
on the (1, 1)-dimensional (anti)-chiral supermanifolds as
\[L_b\longrightarrow \tilde L_b^{(b)} = p_\mu(\tau) \,\dot X^{\mu(b)}(\tau, \bar\theta) - \frac {1}{2}\,E^{(b)}(\tau, \bar\theta)\, (p^2(\tau) - m^2)
+  b(\tau)\,\dot E^{(b)} (\tau, \bar\theta)\]
\[+ \frac {1}{2}\,b^2 (\tau)
 - i\, \dot {\bar F}^{(b)}(\tau, \bar\theta)\,\dot F^{(b)}(\tau, \bar\theta),\]
\[L_b\longrightarrow \tilde L_b^{(ab)} = p_\mu(\tau) \,\dot X^{\mu(ab)}(\tau, \theta) - \frac {1}{2}\,E^{(ab)}(\tau, \theta)\, (p^2(\tau) - m^2)
+  b(\tau) \;\dot E^{(ab)} (\tau, \theta)\]
\[+ \frac {1}{2}\,b^2(\tau)
 - i\, \dot {\bar F}^{(ab)}(\tau, \theta)\,\dot F^{(ab)}(\tau, \theta),\eqno (C.10)\]
where we have taken into account the (anti-)BRST invariance (i.e. $s_{(a)b}\,p_\mu = 0, s_{(a)b}\,b = 0$)
of the variables $p_\mu (\tau)$ and $b(\tau)$ so that $\tilde B^{(b)}(\tau, \bar\theta)  = b(\tau), P_\mu^{(b)}(\tau) = p_\mu(\tau),
\tilde B^{(ab)}(\tau, \bar\theta)  = b(\tau), P_\mu^{(b)}(\tau) = p_\mu(\tau)$
are the $trivial$ expansions. It is elementary  to check that
\[\frac{\partial}{\partial\bar\theta}\,\tilde L_0 =
 \frac {d}{d \tau}\, \Big[\frac {1}{2}\,    c\,(p^2 + m^2) + b\,\dot c\Big]\qquad\Longleftrightarrow\qquad s_b\, L_B,\]
\[\frac{\partial}{\partial\theta}\,\tilde L_0 =
 \frac {d}{d \tau}\, \Big[ \frac {1}{2}\,    \bar c\,(p^2+m^2) + b\,\dot {\bar c}\Big]\qquad\Longleftrightarrow\qquad s_{ab}L_B,\eqno (C.11)\]
which shows the (anti-)BRST invariance of the action integral $S = \int\,d\tau\,L_b$ within the framework of (anti-)chiral
supervariable approach to BRST formalism. In other words, we have captured the (anti-)BRST invariance (cf. Eq. (3)) of Lagrangian $L_b$ 
in the terminology of (anti-)chiral supervariable approach to BRST formalism in the above Eq. (C.11).

We would like to offer a comment here on the above equation (C.11).
Precisely speaking, we should take the derivative w.r.t. $\tau $, on the
r.h.s. of Eq. (C.11), as the {\it partial} derivative because $\tau$ is a part of the superspace coordinates ($\tau, \bar\theta $) and
($\tau, \theta $) that characterize the appropriately chosen (1, 1)-dimensional (anti-)chiral supermanifolds on which the supervariables 
are defined. It is the latter entities that have been taken 
into consideration for our discussion on the (anti-)chiral supervariable approach to BRST formalism.


\begin{thebibliography}{99} 
\bibitem{TM}      J. Thierry-Mieg, {\it J. Math. Phys.} {\bf 21}, 2834 (1980)
\bibitem{MQ}      M. Quiros, F. J. De Urries, J. Hoyos, M. L. Mazon, E. Rodrigues,\\
                  {\it J. Math. Phys.} {\bf 22}, 1767 (1981)              
\bibitem{RPM11}   R. Delbourgo, P. D. Jarvis, {\it J. Phys. A: Math. Gen.} {\bf 15}, 611 (1981)
\bibitem{BT}      L. Bonora, M. Tonin, {\it Phys. Lett.} B {\bf 98}, 48 (1981) 
\bibitem{BPT1}    L. Bonora,  P. Pasti,  M. Tonin, {\it Nuovo Cimento} A {\bf 64}, 307 (1981)
\bibitem{BPT2}    L. Bonora,  P. Pasti, M. Tonin,  {\it Annals of Physics} {\bf 144}, 15 (1982)
\bibitem{RPM32}   L. Baulieu, J. Thierry-Mieg, {\it Nucl. Phys.} B {\bf 197}, 477 (1982)
\bibitem{RPM41}   L. Alvarez-Gaume, L. Baulieu, {\it Nucl. Phys.} B {\bf 212}, 255 (1983)
\bibitem{RPM 15}  G. Curci, R. Ferrari, {\it Phys. Lett.}  B  {\bf 63}, 91 (1976)
\bibitem {RPM 17} L. Bonora, R. P. Malik, {\it Phys. Lett.} B  {\bf 655}, 75  (2007)
\bibitem {RPM 16} L. Bonora, R. P. Malik, {\it J.Phys.} A {\bf 43}, 375403 (2010)
\bibitem{RPM2}    See, e.g., R. P. Malik, {\it J. Phys.  A: Math. Gen.} {\bf 37}, 5261 (2004)
\bibitem{RPM3}    See, e.g., R. P. Malik,  {\it J. Phys. A: Math. Theor.} {\bf 39}, 10575 (2006)
\bibitem{RPM4}    See, e.g., R. P. Malik, {\it Eur. Phys. J.} C {\bf 51}, 169 (2007)
\bibitem{RPM5}    See, e.g.,  R. P. Malik, {\it Eur. Phys. J.} C {\bf 60}, 457 (2009)
\bibitem{RPM6}    See, e.g., A. Shukla, S. Krishna, R. P. Malik \\{\it Advances in High Energy Physics}
                  {\bf 2015}, 258536 (2015) 
\bibitem{RPM12}   N. Srinivas, T. Bhanja, R. P. Malik, arXiv: 1504.04237 [hep-th]
\bibitem{rem 58}  T. Bhanja, N. Srinivas, R. P. Malik \\{\it Advances in High Energy Physics}
                  {\bf 2016}, 3673206 (2016)     
\bibitem{RPM31}   N. Srinivas, T. Bhanja, R. P. Malik,\\
                  {\it Advances in High Energy Physics} {\bf 2017}, 6138263 (2017)
\bibitem{rem 70}      A. Shukla, N. Srinivas, R. P. Malik, {\it Annals of Physics} {\bf 394}, 98  (2018)
                  
\bibitem{rem 72}      B. Chauhan, S. Kumar, R. P. Malik, {\it Int. J. Mod. Phys.} A {\bf 33}, 1850026 (2018) 
\bibitem{rem 71}      S. Kumar, B. Chauhan, R. P. Malik, arXiv: 1712.05242 [hep-th]
\bibitem {REM 21} A. Shukla, S. Krishna, R. P. Malik, {\it Eur. Phys. J.} C  {\bf 72}, 2188 (2012)
\bibitem {REM 22} See, e.g., D. Nemschansky, C. Preitschopf,  M. Weinstein,\\
                  {\it Annals of Physics } {\bf 183}, 226 (1988)
\bibitem {Rem 23}  R. P. Malik, {\it Mod. Phys. Lett.} A {\bf 20}, 1767 (2005)
\bibitem{rem 58}  P. A. M. Dirac, {\it Lectures on Quantum Mechanics}, \\(Yeshiva University Press,
                  New York, 1964)
\bibitem{rem 59}  K. Sundermeyer, {\it Constrained Dynamics: Lecture Notes in Physics} {\bf Vol. 169},\\ (Springer, New
                  York, 1982)
\bibitem {rpm 87} N. Nakanishi, I. Ojima, {\it Covariant Operator Formalism of Gauge Theories\\ and Quantum Gravity} 
                 (World Scientific, Singapore, 1990) 
\bibitem {RPm 14} S. Krishna, A. Shukla, R. P. Malik,  {\it Int. J. Mod. Phys.} A {\bf 29}, 1450183 (2014)
\bibitem {RPm 15} S. Krishna, R. P. Malik, {\it Annals of Physics} {\bf 355}, 204 (2015)
\bibitem {RPm 13} S. Krishna, R. P. Malik, {\it Europhys. Lett.} {\bf 109}, 31001 (2015)
\bibitem {REM 23} S. Krishna, D. Shukla, R. P. Malik, {\it Int. J. Mod. Phys.} A {\bf 31}, 1650113 (2016) 
\bibitem {rem 99} R. P. Malik, {\it etal.}, in preparation               
                  
\end{thebibliography}
\end{document}